\documentclass[aps,prd,twocolumn,nofootinbib,superscriptaddress]{revtex4-2}

\usepackage[utf8]{inputenc}
\usepackage{bm}
\usepackage{amssymb}
\usepackage{amsmath}
\usepackage{graphicx}
\usepackage{tabularx}
\usepackage{enumitem}
\usepackage{color}
\usepackage{hyperref}
\usepackage{booktabs}
\usepackage{array}
\usepackage{upgreek}
\usepackage{appendix}

\usepackage{longtable}

\makeatletter
\renewcommand\appendix{
	\par
	\setcounter{section}{0}%
	\setcounter{subsection}{0}%
	\setcounter{subsubsection}{0}%
	\setcounter{figure}{0}%
	\setcounter{table}{0}%
	\renewcommand{\thesection}{Appendix \Alph{section}}%
	\renewcommand{\thefigure}{\Alph{section}\arabic{figure}}%
	\renewcommand{\thetable}{\Alph{section}\arabic{table}}%	
}
\makeatother

\hypersetup{
	colorlinks = true,
	linkcolor = blue,
	citecolor = blue,
	filecolor = blue,
	urlcolor = blue
}

\def\apj{ApJ}
\def\apjl{ApJL}

\def\aap{Astron. Astrophys.}

\def\mnras{MNRAS}
\def\nat{Nature}

\def\prl{Phys. Rev. Lett.}

\newcolumntype{K}[1]{>{\centering\arraybackslash}p{#1}}

\let\OLDthebibliography\thebibliography
\renewcommand\thebibliography[1]{
	\OLDthebibliography{#1}
	\setlength{\parskip}{0pt}
	\setlength{\itemsep}{0pt plus 0.3ex}
}

\allowdisplaybreaks

\begin{document}
	
	\title{Evidence for an accretion-driven subpopulation of black holes in GWTC-5.0}
	
	\author{Zacharias Roupas}
	%	\email{Zacharias.Roupas@unimib.it}
	\affiliation{Dipartimento di Fisica ``G. Occhialini'', 
		Universit\'a degli Studi di Milano-Bicocca, Piazza della Scienza 3, 20126 Milano, Italy}
	\affiliation{Istituto Nazionale di Fisica Nucleare (INFN), Sezione di Milano-Bicocca, 
		Piazza della Scienza 3, 20126 Milano, Italy}
	
	%	\date{}
	
	\begin{abstract}
	We report Bayesian evidence for a subpopulation of black holes, grown via accretion during the gaseous birth-stage of star clusters, in the cumulative GWTC-5.0 released by the LIGO-Virgo-KAGRA collaboration. This accretion channel predicts a saturating spin--mass relation, with low spins at low masses, rising continuously with mass through all intermediate spin values, to high spins at high masses. We test this prediction directly against the component spins of the catalogue. We single out 10 events with positive support and many others with weaker support across the whole component-mass range of the catalogue.  We identify a preferred mass scale $m^t=20.7^{+11.6}_{-1.2}\,M_\odot$ (90\% credibility) separating regimes with different accretion-mixture fractions, with $\ln B=5.5$. Above this inferred scale, the evidence for an accretion-driven subpopulation is strongly concentrated, reaching a natural log Bayes factor of $16.6$ against the LVK fiducial spin population.
	\end{abstract}
	
	\maketitle
	
	\section{Introduction}
	
	The release of version 5.0 of the cumulative Gravitational-Wave Transient Catalog, GWTC-5.0 \citep{2026arXiv260527225T}, which includes data through the second part of the fourth observing run (O4b), raises the number of confidently detected binary black hole (BBH) mergers to $259$ under a conservative selection \citep{2026arXiv260527226T}.
	Several population-level studies of previous catalogues consistently reported evidence of subpopulations of merging BBHs with distinct mass, effective spin, and mass-ratio properties, associated mostly with the hierarchical-merger channel \citep{2025PhRvL.134a1401A, 2026PhRvL.137b1404P,2026Natur.652..874T}.
	The recent population analysis by the LIGO-Virgo-KAGRA Collaboration (LVK) in GWTC-5.0 \citep{2026arXiv260527226T} provides evidence for subpopulations of BBHs hosting a rapidly spinning black hole (BH) at two scales of primary-component masses $m_1 \sim 10-20\,{\rm M}_{\odot}$ and $m_1 \gtrsim 45\,{\rm M}_{\odot}$. 
	The LVK further infers that at least $9\%$ of mergers occur in channels with some preference for spin-orbit alignment. 
	A non-parametric reconstruction of the joint $m_1$-$\chi_{\rm eff}$ distribution finds again a high-mass subpopulation above a similar scale $\sim 50\,{\rm M}_{\odot}$, which in this case shows a preference for spin-orbit alignment \citep{2026arXiv260623305R}. 
	In GWTC-4 a phenomenological analysis identifies three subpopulations separated by sharp transitions at $27.7^{+4.1}_{-3.4}\,{\rm M}_{\odot}$ and $40.2^{+4.7}_{-3.2}\,{\rm M}_{\odot}$, distinguished by their mass-ratio and spin-magnitude distributions \citep{2026PhRvL.137b1403B}.
	Further evidence of three subpopulations is found using parameterized mixture models \citep{2026ApJ..1005L..55R}. 
Notably, evidence is reported for two subpopulations of BBHs with both BH components spinning, one subpopulation with low spins $\chi \sim 0.1$ and the other with high spins $\chi \sim 0.8$ \citep{2025ApJ...994..261A}.
  	A search for an accretion-origin subpopulation using spin magnitudes as the diagnostic finds a component with primary spins clustered near $\chi\sim 0.9$, which the authors attribute to accretion in active galactic nuclei (AGN) discs \citep{2026arXiv260509351B}.
 Another component-spin analysis reports a BBH subpopulation with both BH components highly spinning whose fraction rises in the range $\sim 35-70\,{\rm M}_{\odot}$ \citep{2026arXiv260524281H}, recovering the same high-mass scale once more. 
	Remarkably, two mass scales, $\sim 20\,{\rm M}_{\odot}$ and $\sim 45\,{\rm M}_{\odot}$, recur among different methods and different observables in recent GW-data population analyses,  the latter attributed to the pair-instability BH mass gap \citep{2026Natur.652..874T}.
	
	We recently proposed an accretion-driven channel of BH growth \citep{2025A&A...702A.208R} which generates a BH spin-mass correlation \citep{2026A&A...709A.5R,2026arXiv260712465R}. 
	It is manifested as a transition from low spins $\chi \lesssim 0.1$ at BH masses $m\lesssim 20\,{\rm M}_{\odot}$ to saturated high spins $\chi \gtrsim 0.8$ at high masses $m \gtrsim 70\,{\rm M}_{\odot}$, with a transition of intermediate spins at the characteristic mass scale, $m\sim 45\,{\rm M}_\odot$. Those mass scales are close to those recovered independently by GW-data analyses. In our proposed channel, the BH growth and spin-mass correlation are an integral part of the formation process of sufficiently massive and dense star clusters---such as those observed at high redshift with the James Webb Space Telescope \citep{2024Natur.632..513A}---especially at subsolar $\sim 0.1\, {\rm Z}_{\odot}$ and low $\lesssim 0.01\, {\rm Z}_{\odot}$ metallicities. The BHs, being the remnants of massive stars, are generated early in the life of the cluster when residual gas from the first star formation event is still present. Lower gas metallicities imply weaker stellar winds which may provide a sufficient time-window for the early-formed BHs to accrete gas before it gets depleted by winds and supernova explosions, if the mass and density of this gaseous proto-cluster allow.
	Since the gas-depletion timescale is relatively short, $\sim 10\,{\rm Myr}$ \citep{2025A&A...702A.208R}, the BHs may accrete gas individually, before they get assembled to BBHs. The component spins therefore need not be aligned with the orbital angular momentum of the binary, though some alignment may also arise dynamically at a later evolutionary stage of the star cluster \citep{2026arXiv260717869R}.
	Here, based on the spin-mass correlation alone, we investigate whether a subpopulation of BHs generated by our accretion-driven channel in gaseous star clusters may be identified in the GWTC-5.0 data.
	
	\section{Simulated clusters}
	
\begin{figure}
		\includegraphics[width=\columnwidth]{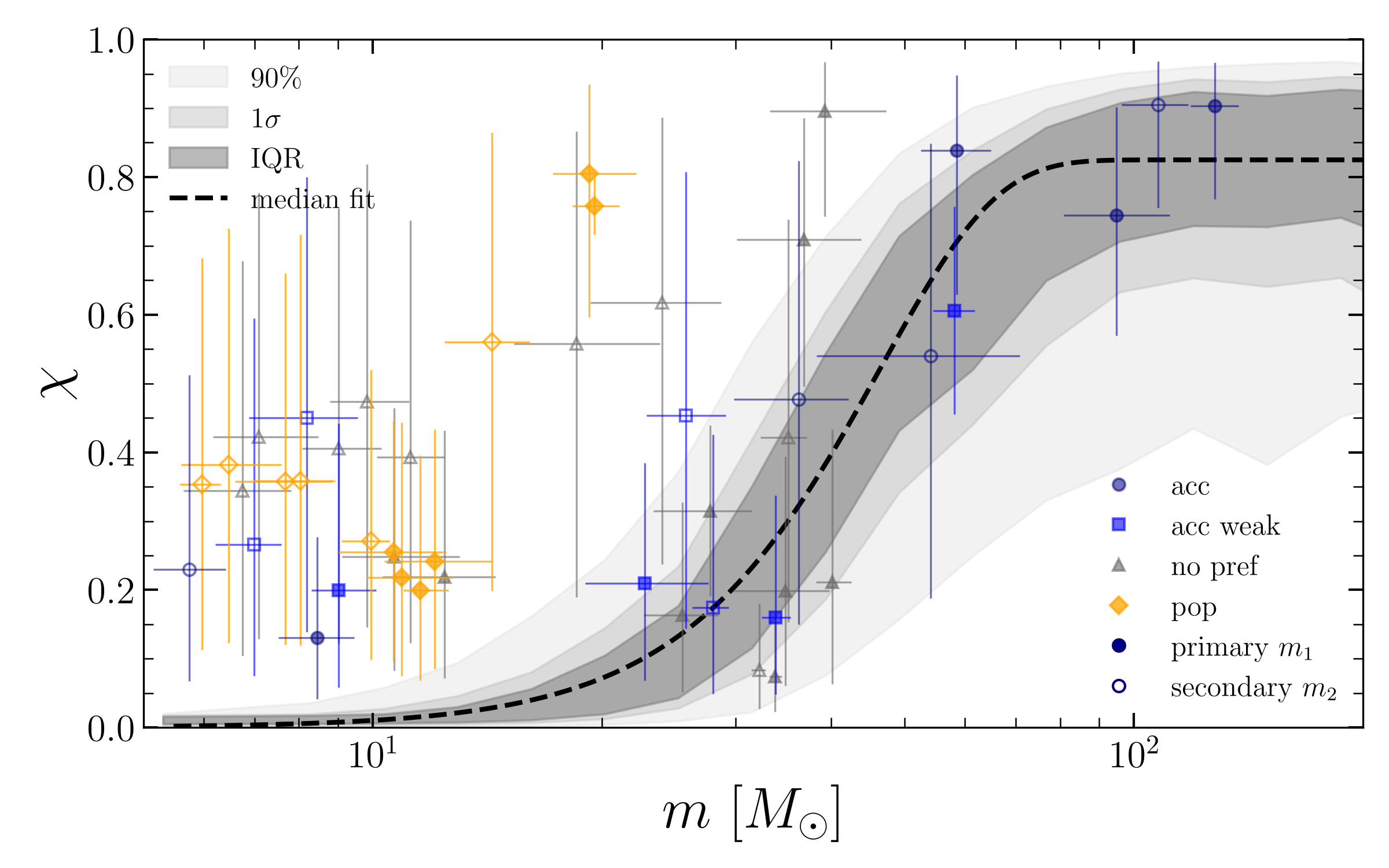}
		\caption{%
			Component spins $\chi$ of LVK binary black holes in the cumulative GWTC-5.0---restricted to events with at least one component with $\Delta_{90}\chi < 0.63$---with respect to their mass $m$,
			against the spin--mass relation predicted by the accretion channel.
			Points: LVK posterior medians with $90\%$ credible intervals, filled markers for
			the primary and open for the secondary; classes as in
			Table~\ref{tab:info_events}. Bands and dashed curve: percentiles and
			median fit Eq.~(\ref{eq:med_fit}) of the predicted $p_{\rm acc}(\chi\mid m)$ by simulations.}
		\label{fig:spinmass}
	\end{figure}

	Our semi-analytic model, developed in Refs.~\citep{2025A&A...702A.208R,2026A&A...709A.5R}, follows the BH mass growth and spin evolution via accretion within a gaseous star cluster for $\sim 10\,{\rm Myr}$, during which the residual gas from the first star-formation event gets progressively depleted by $99\%$.
	A persistent spin-mass correlation of BHs is generated for sufficiently massive and dense clusters with
	a median spin which can be consistently fitted with a saturated exponential law of the form
	\begin{equation}\label{eq:med_fit}
		\bar{\chi}_{\rm acc}(m) = \chi_{\rm max}\left(1 - e^{-\left(\frac{m}{m_{\rm T}}\right)^{\beta}}\right), 
	\end{equation}
	where the parameters vary only slightly. 
	For the various star-cluster configurations we considered, described in \ref{app:clusters}, 
	the maximum spin gets values $\chi_{\rm max} = 0.74-0.90$, the characteristic mass scale $m_{\rm T} = 40-55\,{\rm M}_{\odot}$, and the exponent $\beta = 2.5-3.3$.
	For the pooled sample of clusters we get the fit $\{\chi_{\rm max}, m_{\rm T}, \beta\} = \{0.83, 46.5\,{\rm M}_{\odot}, 2.85\}$, displayed as a dashed line in Figure~\ref{fig:spinmass}. 
	We discuss in more detail the assumptions regarding our cluster modelling in \ref{app:clusters}.
	
	The Eq.~(\ref{eq:med_fit}) demonstrates the consistency of the form of the spin-mass correlation in our accretion-driven channel and is used for illustrative purposes in Figure~\ref{fig:spinmass}, along with percentiles. However, significant outliers, especially in the low-mass/intermediate-spin band (see Figure \ref{fig:a_m_fin_data}), arise outside the $90\%$ credible range. Therefore, we calculate the simulated conditional density $p_{\rm acc}(\chi | m)$ directly from the simulation data as a piecewise function along $(\chi, m)$ cells. The probability in the $(b,c)$ cell is then
	\begin{equation}\label{eq:pacc}
		p_{{\rm acc},(b,c)} =
		\frac{1}{\Delta\chi_{c}}
		\frac{N_{(b,c)}}{N_{b}},
	\end{equation}
	where $N_{(b,c)}$ denotes the number of simulated BHs in the cell, $N_{b}$ the total number of simulated BHs in the $b$ mass bin, and $\Delta\chi_{c}$ the $\chi$ range of the cell.
	The pooled probability density then sums over all weighted clusters as discussed in \ref{app:clusters}.
	
	\begin{table*}
		\caption{%
			BBH mergers in the LVK cumulative GWTC-5.0 with at least one informative
			component-spin measurement, $\Delta_{90}\chi < 0.64$.
			$m_1,m_2$: medians of source-frame component masses $[{\rm M_\odot}]$;
			$\chi_1,\chi_2$: medians of component spin magnitudes;
			$\chi_{\rm eff}$: median effective inspiral spin;
			$\Delta_{90}\chi_i$: width of the $90\%$ credible interval of $\chi_i$;
			$\ln B^{\rm Def}$, $\ln B^{\rm Merg}$: log Bayes factor of accretion
			against the \textsc{Default-bbh} model and  against the hierarchical-merger channel \textsc{Merg}, with their sampling
			uncertainties; $P^{\rm Def}_{\rm acc}$: accretion membership probability;
			Class: classification flag based on the  evidence.}
		\label{tab:info_events}
		\squeezetable
		\begin{ruledtabular}
			\begin{tabular}{lrrrrrrrrrrl}
				Event & $m_1$ & $m_2$ & $\chi_1$ & $\chi_2$ & $\chi_{\rm eff}$ &
				$\Delta_{90}\chi_1$ & $\Delta_{90}\chi_2$ &
				$\ln B^{\rm Def}$ & $\ln B^{\rm Merg}$ & $P^{\rm Def}_{\rm acc}$ & Class 
				\\
				\midrule
				GW231123\_135430 & $127.84$ & $107.70$ & $0.903$ & $0.905$ & $0.270$ & $0.355$ & $0.394$ & $+4.782\pm0.018$ & $+4.473\pm0.017$ & $0.9850$ & \mbox{acc} \\
				GW230627\_015337 & $8.45$ & $5.74$ & $0.131$ & $0.230$ & $0.023$ & $0.388$ & $0.707$ & $+1.647\pm0.063$ & $+1.653\pm0.063$ & $0.7498$ & \mbox{acc} \\
				GW231028\_153006 & $94.85$ & $54.11$ & $0.744$ & $0.540$ & $0.388$ & $0.511$ & $0.883$ & $+1.625\pm0.016$ & $+1.449\pm0.015$ & $0.7459$ & \mbox{acc} \\
				GW241225\_082815 & $58.53$ & $36.27$ & $0.838$ & $0.477$ & $-0.225$ & $0.551$ & $0.893$ & $+1.357\pm0.020$ & $+1.208\pm0.020$ & $0.6939$ & \mbox{acc} \\
				GW240925\_005809 & $9.02$ & $6.99$ & $0.199$ & $0.266$ & $0.025$ & $0.615$ & $0.783$ & $+0.978\pm0.073$ & $+0.988\pm0.073$ & $0.6116$ & \mbox{acc weak} \\
				GW241127\_061008 & $58.15$ & $25.80$ & $0.606$ & $0.454$ & $-0.134$ & $0.489$ & $0.885$ & $+0.820\pm0.020$ & $+0.744\pm0.020$ & $0.5753$ & \mbox{acc weak} \\
				GW231114\_043211 & $22.76$ & $8.19$ & $0.210$ & $0.450$ & $0.081$ & $0.485$ & $0.883$ & $+0.584\pm0.051$ & $+0.631\pm0.051$ & $0.5193$ & \mbox{acc weak} \\
				GW230814\_230901 & $33.84$ & $27.98$ & $0.160$ & $0.174$ & $-0.006$ & $0.447$ & $0.608$ & $+0.527\pm0.014$ & $+0.616\pm0.014$ & $0.5056$ & \mbox{acc weak} \\
				GW250114\_082203 & $33.77$ & $32.21$ & $0.075$ & $0.084$ & $-0.023$ & $0.246$ & $0.263$ & $+0.492\pm0.011$ & $+0.593\pm0.011$ & $0.4972$ & \mbox{no pref} \\
				GW240930\_035959 & $25.53$ & $11.21$ & $0.163$ & $0.393$ & $-0.051$ & $0.443$ & $0.856$ & $+0.486\pm0.048$ & $+0.546\pm0.048$ & $0.4956$ & \mbox{no pref} \\
				GW240515\_005301 & $36.85$ & $18.51$ & $0.709$ & $0.558$ & $0.449$ & $0.623$ & $0.890$ & $+0.174\pm0.056$ & $+0.027\pm0.056$ & $0.4215$ & \mbox{no pref} \\
				GW241231\_054133 & $12.42$ & $7.09$ & $0.219$ & $0.422$ & $0.084$ & $0.577$ & $0.879$ & $+0.138\pm0.186$ & $+0.177\pm0.186$ & $0.4130$ & \mbox{no pref} \\
				GW191129\_134029 & $10.68$ & $6.75$ & $0.248$ & $0.344$ & $0.064$ & $0.595$ & $0.827$ & $+0.011\pm0.160$ & $+0.038\pm0.160$ & $0.3839$ & \mbox{no pref} \\
				GW190412\_053044 & $27.75$ & $9.02$ & $0.315$ & $0.406$ & $0.214$ & $0.411$ & $0.869$ & $-0.039\pm0.065$ & $+0.009\pm0.065$ & $0.3725$ & \mbox{no pref} \\
				GW190517\_055101 & $39.24$ & $24.02$ & $0.896$ & $0.617$ & $0.491$ & $0.394$ & $0.885$ & $-0.079\pm0.078$ & $-0.287\pm0.078$ & $0.3636$ & \mbox{no pref} \\
				GW240921\_201835 & $34.86$ & $9.83$ & $0.199$ & $0.474$ & $0.005$ & $0.541$ & $0.889$ & $-0.098\pm0.048$ & $-0.053\pm0.048$ & $0.3593$ & \mbox{no pref} \\
				GW231226\_101520 & $40.16$ & $35.15$ & $0.211$ & $0.422$ & $-0.065$ & $0.593$ & $0.845$ & $-0.253\pm0.016$ & $-0.208\pm0.016$ & $0.3257$ & \mbox{no pref} \\
				GW241102\_124058 & $10.91$ & $8.04$ & $0.218$ & $0.359$ & $0.064$ & $0.611$ & $0.850$ & $-2.206\pm0.152$ & $-2.168\pm0.152$ & $0.0667$ & \mbox{pop} \\
				GW250119\_190238 & $11.54$ & $9.95$ & $0.199$ & $0.271$ & $0.089$ & $0.572$ & $0.708$ & $-2.647\pm0.035$ & $-2.575\pm0.035$ & $0.0442$ & \mbox{pop} \\
				GW191216\_213338 & $12.07$ & $7.67$ & $0.241$ & $0.358$ & $0.111$ & $0.577$ & $0.823$ & $-2.991\pm0.081$ & $-2.951\pm0.081$ & $0.0318$ & \mbox{pop} \\
				GW240910\_103535 & $10.65$ & $6.47$ & $0.255$ & $0.382$ & $0.108$ & $0.572$ & $0.857$ & $-3.162\pm0.084$ & $-3.137\pm0.084$ & $0.0269$ & \mbox{pop} \\
				GW241113\_163507 & $19.26$ & $14.36$ & $0.805$ & $0.560$ & $0.496$ & $0.551$ & $0.883$ & $-3.774\pm0.624$ & $-3.996\pm0.624$ & $0.0148$ & \mbox{pop} \\
				GW241011\_233834 & $19.54$ & $5.96$ & $0.758$ & $0.354$ & $0.504$ & $0.138$ & $0.841$ & $-4.899\pm0.077$ & $-5.175\pm0.077$ & $0.0049$ & \mbox{pop} \\
			\end{tabular}
		\end{ruledtabular}
	\end{table*}
	
	\section{GWTC-5.0 population-informed priors} 
	We wish to investigate the evidence for the accretion-driven density in the LVK GWTC-5.0 events. We select waveform models per event following the LVK prescription for each catalog \citep{2026ApJ..1005L..51A,2026arXiv260527226T}.
	We define the Bayes factor per event
		\begin{equation}
		B^X =
		\frac{\left\langle p_{\rm acc}(\bm{\chi}\mid\bm{m})/
			\pi_{\rm PE}(\bm{\chi})\right\rangle}
		{\left\langle p_{X}(\bm{\chi}\mid\bm{m})/
			\pi_{\rm PE}(\bm{\chi})\right\rangle},
		\label{eq:bayes}
	\end{equation}	
	where $\pi_{\rm PE}$ is the parameter-estimation prior, and
	both averages are taken over the same joint posterior samples of the event.
	Since no joint spin prior is released, we take 
	$\pi_{\rm PE}(\bm{\chi}) = \pi_{\rm PE}(\chi_1) \pi_{\rm PE}(\chi_2)$, where each factor is built from that event's released prior samples for each component, or from the prior of the corresponding catalogue when these are not released. 
	The component-spin prior is uniform for all events with released samples but one.
	We denote $p_X$ the population-informed prior on the component spins of
	the event. 
	
	We compare the accretion prediction with three reference models. 
	Our null is the LVK \textsc{Default-bbh} model
	\citep{2026arXiv260527226T}, the fiducial description of the whole BBH population over GWTC-5.0, which draws both spins from a common truncated Gaussian, $\Lambda=(\mu_\chi,\sigma_\chi)$, and carries no mass dependence. 
	The second comparative model, we call \textsc{2br}, is also a GWTC-5.0 population-informed model, which already asserts a rapidly spinning subpopulation.
	The third comparative model, denoted \textsc{Merg}, is designed to capture the characteristic spin distribution of the hierarchical-merger channel.
	
	For each hypothesis we use the spin-magnitude sector alone.
	We marginalise over its uncertain hyperparameters,
	\begin{equation}\label{eq:ppd}
		p_X(\bm{\chi}\mid\bm{m})
		= \int d\Lambda\,\Pi_X(\Lambda)\,
		p_X(\bm{\chi}\mid\bm{m},\Lambda),
	\end{equation}
	where $\Pi_X$ denotes the adopted hyperparameter distribution.
	For \textsc{Default-bbh} and \textsc{2br}, it is $\Pi_X= p(\Lambda|D)$, the LVK hyperposterior inferred from GWTC-5.0 itself. These comparison distributions are therefore population-informed posterior predictives.
	For \textsc{Merg}, the LVK hyperposterior supplies the
	first-generation spin distribution, while the second-generation fraction is marginalised over the explicit prior specified below.
	
	The \textsc{2br} hypothesis is the spin sector of the Three
	Subpopulation model of \citet{2026PhRvL.137b1403B}, which
	\citet{2026arXiv260527226T} refit to GWTC-5.0 and use to constrain the
	onset of a high-mass subpopulation to
	$m^t_{\rm 2br} = 46.6^{+11.6}_{-6.7}\,{\rm M}_{\odot}$. The spin sector
	has two branches, hence our label; the mass-ratio sector does not
	enter. The Gaussian switches between them at $m^t_{\rm 2br}$,
	\begin{equation}\label{eq:2br}
		p_{\rm 2br}(\bm{\chi}\mid\bm{m},\Lambda) =
		\prod_{i=1,2}\mathcal{N}_{[0,1]}(\chi_i;\mu_b,\sigma_b),
	\end{equation}
	where the branch index $b$ separates $m_1 \leq m^t_{\rm 2br}$ from
	$m_1 > m^t_{\rm 2br}$, and $\Lambda=(m^t_{\rm 2br},\mu_{\rm low},
	\sigma_{\rm low},\mu_{\rm high},\sigma_{\rm high})$ is marginalised
	over its hyperposterior. 
	The inferred onset is consistent with the characteristic scale $m_T=46.5\,M_\odot$ of our pooled simulated clusters. Thus, the \textsc{2br} model already captures phenomenologically one of the principal features predicted by the accretion channel, namely the emergence of a distinct high-mass spin population. 
	The factor $\ln K^{\rm 2br}$ therefore provides a more demanding comparison than $\ln K^{\rm Def}$: it tests whether the accretion-driven spin--mass distribution is preferred over an LVK-informed phenomenological
	description that already accommodates a high-mass spin transition.
Consequently, $\ln K^{\rm 2br}\lesssim0$ would not by itself exclude an accretion contribution; it would imply only that the current data do not distinguish its specific spin--mass structure from this phenomenological two-branch description. Conversely, a positive $\ln K^{\rm 2br}$ provides additional evidence that the data prefer features of the predicted accretion spin--mass distribution beyond the mere existence of a
high-mass spin transition.
	
	The \textsc{Merg} comparison captures the characteristic spin-magnitude signature of hierarchical-merger remnants. It is deliberately not a complete hierarchical-formation model, so that mass-ratio, redshift, and spin-orientation predictions are not included in the Bayes factor considered here. We assume that a component above the second-generation mass floor, which we assume at $m_{\rm 2g} = 9.5\,{\rm M}_{\odot}$, the lightest possible merger remnant, is itself a remnant with
	probability $g$, in which case its spin is drawn from $p_{\rm R} = \mathrm{Beta}(6.6,3.4)$, of mode
	$0.7$ and concentration $\nu=10$, reproducing the near-universal remnant spin of a comparable-mass merger
	\citep{2008ApJ...684..822B,2017ApJ...840L..24F}, otherwise it follows \textsc{Default-bbh}
	\begin{align}\label{eq:merg}
		p_{\rm Merg}&(\bm{\chi}\mid\bm{m},\Lambda) =
		\nonumber \\
		&\prod_{i=1,2}\Bigl[g_i\, p_{\rm R}(\chi_i)
		+ (1-g_i)\,\mathcal{N}_{[0,1]}(\chi_i;\mu_\chi,\sigma_\chi)\Bigr].
	\end{align}
	We denote $g_i = g$ for $m_i \ge m_{\rm 2g}$ and $g_i = 0$ otherwise, and
	$\Lambda = (\mu_\chi,\sigma_\chi,g)$. The same $g$ applies to both components. It is marginalised in Eq.~(\ref{eq:ppd}) over a uniform prior on $[0,g_{\rm max}]$, which integrates in closed form.
	We take $g_{\rm max} = 0.1$, which allows up to $\simeq 19\%$ of binaries above the floor to contain a rapidly spinning component, more than the $15\%$ median inferred by the LVK \citep{2026arXiv260527226T}.

	\begin{figure}
		\includegraphics[width=\columnwidth]{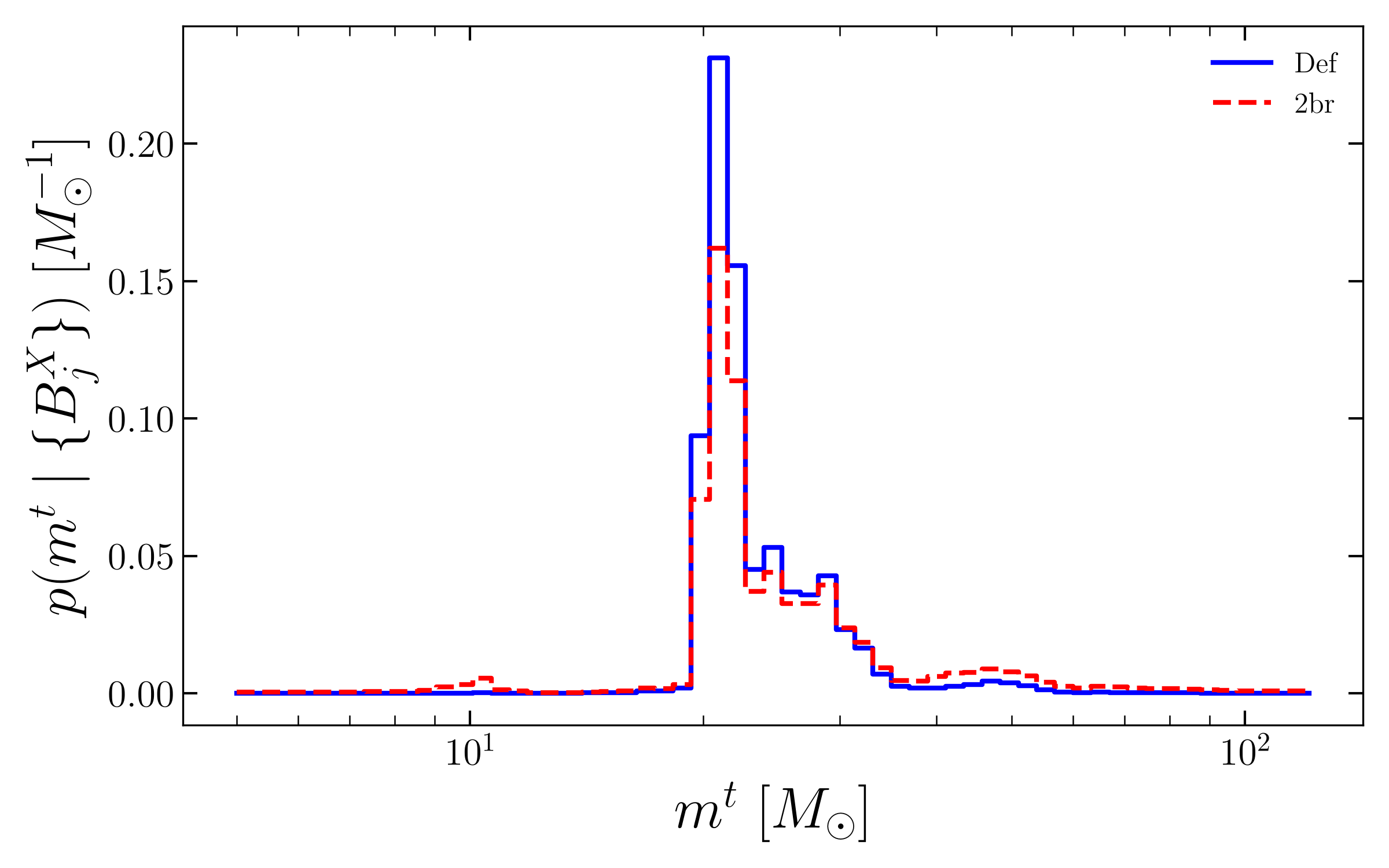}
		\caption{%
			Posterior on the transition mass $m^t$ against the LVK
			\textsc{Default-bbh} and \textsc{2br} comparisons.}
		\label{fig:mass_scale_hist}
	\end{figure}
	
	\begin{table}
		\caption{%
			Mixture evidence by primary-mass span. $N$: number of events in the span;
			$\ln K^{\rm Def}$, $\ln K^{\rm 2br}$, $\ln K^{\rm Merg}$: log mixture Bayes
			factors of the accretion model against the \textsc{Default-bbh}, \textsc{2br} models,
			and the hierarchical-merger channel \textsc{Merg}. The last two rows overlap the disjoint spans
			above; $\ln K$ is not additive across spans.}
		\label{tab:massspan}
		\begin{ruledtabular}
			\begin{tabular}{lrrrr}
				Span $[{\rm M_\odot}]$ & $N$ & $\ln K^{\rm Def}$ & $\ln K^{\rm 2br}$ &
				$\ln K^{\rm Merg}$ \\
				\midrule
				$m_1\ge46$ & $58$ & $13.57$ & $4.69$ & $12.19$ \\
				$20\le m_1<46$ & $136$ & $2.80$ & $1.80$ & $4.15$ \\
				$10\le m_1<20$ & $59$ & $-2.13$ & $-2.29$ & $-2.01$ \\
				$m_1<10$ & $6$ & $3.77$ & $3.48$ & $3.80$ \\
				\midrule
				all masses & $259$ & $8.76$ & $1.22$ & $8.99$ \\
				$m_1\ge20$ & $194$ & $16.59$ & $6.56$ & $16.84$ \\
			\end{tabular}
		\end{ruledtabular}
	\end{table}
	
	\section{Population mixture}
	
	A fraction $f$ of the catalogue is drawn from the accretion channel and
	the rest from the population-informed prior $X$. The evidence for this
	mixture against $f=0$ follows from the per-event Bayes factors of
	Eq.~(\ref{eq:bayes}),
	\begin{equation}\label{eq:mixture}
		K^X = \int_0^1\! df \mathcal{L}^X(f),\;
		\mathcal{L}^X \equiv
		\prod_{j} \left[(1-f) + f B^X_j\right],
	\end{equation}
	where $j$ runs over the $259$ events and the prior on $f$ is uniform.
	The membership probability of event $j$ is the
	responsibility of the accretion component under the same posterior,
	\begin{equation}
		P^X_{{\rm acc},j} = \int_0^1\! df\,
		\frac{f B^{X}_j}{(1-f) + f B^{X}_j}
		\,\frac{\mathcal{L}^{X}(f)}{K^{X}}.
		\label{eq:pacc_event}
	\end{equation}
	Both integrals marginalise $f$ under its uniform prior; we do not fit it. Note that $f$ weights the detected catalogue at fixed observed masses and is not an intrinsic branching fraction, which would additionally require the merger rate and the mass and redshift distributions predicted by the channel, along with the corresponding detection efficiency.
	
	We wish further to investigate whether the accretion channel forms a distinct subpopulation in mass. We determine a transition mass scale $m^t$ by Bayesian model selection between a constant and a change-point mixing fraction. The fraction takes two values across the change-point, $f_j = f_{\rm L}$ for $m_{1,j} < m^t$ and $f_j = f_{\rm H}$ above it, and all three parameters are marginalised,
\begin{equation}\label{eq:K1}
	K_1^X = \int\! dm^t\, \pi(m^t)
	\int_0^1\! df_{\rm L} \int_0^1\! df_{\rm H}\;
	\mathcal{L}^X(f_j),
\end{equation}
where the mass integral runs from the minimum BH mass of the analysis $5\,{\rm M}_{\odot}$ to the heaviest primary in the sample. Every event enters at every $m^t$, so
$\ln B_{10}^X = \ln K_1^X - \ln K^X$ is a Bayes factor measuring the
evidence for a mass-dependent accretion fraction. The posterior on the
change-point is the same integrand normalised
\begin{equation}\label{eq:pmtB}
p(m^t\mid\{B^X_j\}) \propto \pi(m^t)\,K^X_{\rm L}(m^t)\,
K^X_{\rm H}(m^t),
\end{equation}
with $K^X_{\rm L}$ and $K^X_{\rm H}$ the integrals
over $f_{\rm L}$ and $f_{\rm H}$ of the events below and above $m^t$.
The likelihood changes only when an event crosses $m^t$, so the marginalisation is an exact sum over the $N+1$ partitions of the mass-ordered catalogue. We take $\pi(m^t)$ uniform on this range, and
repeat with a log-uniform prior, the scale-invariant alternative over this range. 
We quote only the results that both priors share.

We repeated the procedure with two change-points, a discrete analogue of the mass-dependent mixing fraction of \citet{2026arXiv260524281H}. The second change-point raises the evidence by less than $1.4$ and its
fitted scales move by a factor of two between the two priors, therefore we report only the single change-point throughout.
	
\section{Results}

In Figure \ref{fig:spinmass} we plot the spin-mass scatter diagram of BH components for LVK events with at least one component having $\Delta_{90}\chi < 0.63$, corresponding to half a bit of new information in the spin posterior. The BH data are overlaid by the accretion channel prediction and the median spin fit Eq.~(\ref{eq:med_fit}).
The complete per-event results for the whole catalogue are
given in Table~\ref{tab:full_catalogue}. Of the $259$ events, $10$ show positive evidence $\ln B^{\rm Def}>1$ (`acc') and a further $29$ reach $0.5<\ln B^{\rm Def}\le1$ (`acc weak'), spanning the full
component-mass range of the catalogue. 
This includes $17$ events with $m_1 < 20\,{\rm M}_\odot$ ($6$ positive and $11$ weak), whose component spins, $\chi \simeq 0.13-0.43$, lie above the near-zero $90\%$ credible band of Eq.~(\ref{eq:med_fit}) but within the low-mass/intermediate-spin outlier population of the simulations (see Figure~\ref{fig:a_m_fin_data}), which $p_{\rm acc}(\chi\mid m)$ retains by construction, Eq.~(\ref{eq:pacc}).
Note that only four events---GW230627, GW231028, GW231123, GW241225 082815---of the ten positive ones appear in Figure~\ref{fig:spinmass} (see also Table \ref{tab:info_events}) . The remaining six events---GW190707, GW191109, GW231224, GW240428, GW240629, GW241109\_115924---have $0.65 \leq \Delta_{90}\chi_1 \leq 0.81$ (see Table~\ref{tab:full_catalogue}), yet their joint two-dimensional
posterior still favours the accretion channel over \textsc{Default-bbh}.

We further determine the mass range where the accretion-driven
subpopulation is best localised, allowing the mixing fraction to take independent values below and above a free transition mass scale $m^t$. A mass-dependent accretion fraction is decisively favoured over a constant one: $\ln B_{10}^{\rm Def}=+5.54$ against \textsc{Default-bbh} and $+2.80$ against \textsc{2br}. 
The significance of the transition against a null of no mass dependence, built by permuting Bayes factors across masses, is $p=0.001$ (\textsc{Default-bbh}) and $p=0.002$ (\textsc{2br}) in a thousand permutations. 
The posterior mode, Figure~\ref{fig:mass_scale_hist}, is identical, $m^t=20.74\,{\rm M}_\odot$, for both nulls and under both the uniform and log-uniform priors, with a $90\%$ narrowest credible interval $[19.5,32.3]\,{\rm M}_\odot$ and a $95\%$ lower bound $m^t>20.1\,{\rm M}_\odot$ under \textsc{Default-bbh}; $67\%$ of the posterior lies in $20-26\,{\rm M}_\odot$.

We calculate the evidence in favour of an accretion-driven subpopulation for several primary-mass bands, reported in Table~\ref{tab:massspan}.
The LVK analysis \citep{2026arXiv260527226T} in the GWTC-5.0 suggests that rapidly-spinning subpopulations exist in two separate bands, $10\,{\rm M}_{\odot} \lesssim m_1 \lesssim 20\,{\rm M}_\odot$ and $m_1 \gtrsim 45\,{\rm M}_{\odot}$. Therefore, we included those bands in our results. The upper band is notably consistent with the characteristic mass $m_{\rm T}$ of our predicted median fit Eq.~(\ref{eq:med_fit}). 
The upper edge of the LVK low-mass rapidly-spinning population,
$\sim 20\,{\rm M}_\odot$, lies within the credible region of our inferred change point $m^t$.
The evidence in favour of the accretion channel is strongly concentrated above this representative boundary, $m_1\ge20\,{\rm M}_\odot$, with $\ln K^{\rm Def}=16.6$.

We find that the accretion channel is additionally favoured at low masses $<10\,{\rm M}_{\odot}$ although in a small sample, and for high masses $\geq 46\,{\rm M}_{\odot}$. It is disfavoured against all three comparison models in the range $10-20\,{\rm M}_{\odot}$.

\section{Discussion}
The GWTC-5.0 component-spin measurements favour an accretion-driven subpopulation for $m_1\gtrsim 20\,M_\odot$ relative to two LVK population-informed spin models and to our merger-remnant spin-magnitude comparator, \textsc{Merg}. The LVK comparison distributions were themselves inferred from GWTC-5.0 and are therefore not statistically independent of the events tested here.
The result is carried by the population itself, not by any single event. In particular, excluding the most massive and most strongly favoured merger, GW231123, still leaves strong evidence $\ln K^{\rm Def}=12.0$ in the same mass range. 

We explicitly compare the accretion model over several mass ranges with the characteristic spin-magnitude signature of hierarchical merger remnants, which provides an important alternative explanation for rapidly spinning black holes.
We find that the band $10-20\,{\rm M}_\odot$ disfavours the accretion channel against \textsc{Merg} (as well as the other two comparison models). 
On the other hand, the accretion model is strongly favoured against \textsc{Merg} at high masses, $m_1\ge 46\,{\rm M}_\odot$, with $\ln K^{\rm Merg}=12.2$. 
The two channels differ in the shape and values of the spin distribution they predict. 
Merger remnants are expected to cluster near the characteristic spin $\chi\simeq0.7$, whereas accretion produces a continuous $\chi(m)$ that passes through all intermediate values (Figure~\ref{fig:spinmass}), comfortably reaching extreme spin values $\sim 0.9$ at high masses. 
Therefore, intermediate spins, $\sim 0.1-0.5$, populated at $20-46\,{\rm M}_\odot$, and extreme spins at high masses provide discriminating features between the accretion prediction and the merger-remnant spin comparator. 
In addition, the low-mass band $m_1 < 10\,{\rm M}_{\odot}$ disfavours a second-generation merger-remnant interpretation, assuming a $5\,{\rm M}_{\odot}$ minimum BH mass. Nevertheless, it can provide evidence for the accretion channel, which can still accommodate low-mass/intermediate-spin outliers.

Future catalogues will sharpen the test reported here, especially since the prediction covers the whole mass and spin range. The test will benefit greatly from improved component-spin precision. A key target is the transition region $20-70\,{\rm M}_\odot$, where this accretion channel, in contrast to the merger channel, predicts a continuous rise from low to saturated high spins.

	\section*{Acknowledgments}
	\noindent	
	ZR is supported by the European Union's Horizon Europe Research and Innovation Programme under the Marie Sk\l{}odowska-Curie grant agreement No.~101149270--ProtoBH.
	
	\bibliography{protoBH_population}

%apsrev4-2.bst 2019-01-14 (MD) hand-edited version of apsrev4-1.bst
%Control: key (0)
%Control: author (72) initials jnrlst
%Control: editor formatted (1) identically to author
%Control: production of article title (-1) disabled
%Control: page (0) single
%Control: year (1) truncated
%Control: production of eprint (0) enabled
\begin{thebibliography}{29}%
\makeatletter
\providecommand \@ifxundefined [1]{%
 \@ifx{#1\undefined}
}%
\providecommand \@ifnum [1]{%
 \ifnum #1\expandafter \@firstoftwo
 \else \expandafter \@secondoftwo
 \fi
}%
\providecommand \@ifx [1]{%
 \ifx #1\expandafter \@firstoftwo
 \else \expandafter \@secondoftwo
 \fi
}%
\providecommand \natexlab [1]{#1}%
\providecommand \enquote  [1]{``#1''}%
\providecommand \bibnamefont  [1]{#1}%
\providecommand \bibfnamefont [1]{#1}%
\providecommand \citenamefont [1]{#1}%
\providecommand \href@noop [0]{\@secondoftwo}%
\providecommand \href [0]{\begingroup \@sanitize@url \@href}%
\providecommand \@href[1]{\@@startlink{#1}\@@href}%
\providecommand \@@href[1]{\endgroup#1\@@endlink}%
\providecommand \@sanitize@url [0]{\catcode `\\12\catcode `\$12\catcode
  `\&12\catcode `\#12\catcode `\^12\catcode `\_12\catcode `\%12\relax}%
\providecommand \@@startlink[1]{}%
\providecommand \@@endlink[0]{}%
\providecommand \url  [0]{\begingroup\@sanitize@url \@url }%
\providecommand \@url [1]{\endgroup\@href {#1}{\urlprefix }}%
\providecommand \urlprefix  [0]{URL }%
\providecommand \Eprint [0]{\href }%
\providecommand \doibase [0]{https://doi.org/}%
\providecommand \selectlanguage [0]{\@gobble}%
\providecommand \bibinfo  [0]{\@secondoftwo}%
\providecommand \bibfield  [0]{\@secondoftwo}%
\providecommand \translation [1]{[#1]}%
\providecommand \BibitemOpen [0]{}%
\providecommand \bibitemStop [0]{}%
\providecommand \bibitemNoStop [0]{.\EOS\space}%
\providecommand \EOS [0]{\spacefactor3000\relax}%
\providecommand \BibitemShut  [1]{\csname bibitem#1\endcsname}%
\let\auto@bib@innerbib\@empty
%</preamble>
\bibitem [{\citenamefont {{The LIGO Scientific Collaboration}}\ \emph
  {et~al.}(2026{\natexlab{a}})\citenamefont {{The LIGO Scientific
  Collaboration}}, \citenamefont {{the Virgo Collaboration}}, \citenamefont
  {{the KAGRA Collaboration}} \emph {et~al.}}]{2026arXiv260527225T}%
  \BibitemOpen
  \bibfield  {author} {\bibinfo {author} {\bibnamefont {{The LIGO Scientific
  Collaboration}}}, \bibinfo {author} {\bibnamefont {{the Virgo
  Collaboration}}}, \bibinfo {author} {\bibnamefont {{the KAGRA
  Collaboration}}}, \emph {et~al.},\ }\href
  {https://doi.org/10.48550/arXiv.2605.27225} {\bibfield  {journal} {\bibinfo
  {journal} {arXiv e-prints}\ ,\ \bibinfo {eid} {arXiv:2605.27225}} (\bibinfo
  {year} {2026}{\natexlab{a}})},\ \Eprint {https://arxiv.org/abs/2605.27225}
  {arXiv:2605.27225 [gr-qc]} \BibitemShut {NoStop}%
\bibitem [{\citenamefont {{The LIGO Scientific Collaboration}}\ \emph
  {et~al.}(2026{\natexlab{b}})\citenamefont {{The LIGO Scientific
  Collaboration}}, \citenamefont {{the Virgo Collaboration}}, \citenamefont
  {{the KAGRA Collaboration}} \emph {et~al.}}]{2026arXiv260527226T}%
  \BibitemOpen
  \bibfield  {author} {\bibinfo {author} {\bibnamefont {{The LIGO Scientific
  Collaboration}}}, \bibinfo {author} {\bibnamefont {{the Virgo
  Collaboration}}}, \bibinfo {author} {\bibnamefont {{the KAGRA
  Collaboration}}}, \emph {et~al.},\ }\href
  {https://doi.org/10.48550/arXiv.2605.27226} {\bibfield  {journal} {\bibinfo
  {journal} {arXiv e-prints}\ ,\ \bibinfo {eid} {arXiv:2605.27226}} (\bibinfo
  {year} {2026}{\natexlab{b}})},\ \Eprint {https://arxiv.org/abs/2605.27226}
  {arXiv:2605.27226 [astro-ph.HE]} \BibitemShut {NoStop}%
\bibitem [{\citenamefont {{Antonini}}\ \emph {et~al.}(2025)\citenamefont
  {{Antonini}}, \citenamefont {{Romero-Shaw}},\ and\ \citenamefont
  {{Callister}}}]{2025PhRvL.134a1401A}%
  \BibitemOpen
  \bibfield  {author} {\bibinfo {author} {\bibfnamefont {F.}~\bibnamefont
  {{Antonini}}}, \bibinfo {author} {\bibfnamefont {I.~M.}\ \bibnamefont
  {{Romero-Shaw}}},\ and\ \bibinfo {author} {\bibfnamefont {T.}~\bibnamefont
  {{Callister}}},\ }\href {https://doi.org/10.1103/PhysRevLett.134.011401}
  {\bibfield  {journal} {\bibinfo  {journal} {\prl}\ }\textbf {\bibinfo
  {volume} {134}},\ \bibinfo {eid} {011401} (\bibinfo {year} {2025})},\ \Eprint
  {https://arxiv.org/abs/2406.19044} {arXiv:2406.19044 [astro-ph.HE]}
  \BibitemShut {NoStop}%
\bibitem [{\citenamefont {{Plunkett}}\ \emph {et~al.}(2026)\citenamefont
  {{Plunkett}} \emph {et~al.}}]{2026PhRvL.137b1404P}%
  \BibitemOpen
  \bibfield  {author} {\bibinfo {author} {\bibfnamefont {C.}~\bibnamefont
  {{Plunkett}}} \emph {et~al.},\ }\href {https://doi.org/10.1103/n6p4-ftgq}
  {\bibfield  {journal} {\bibinfo  {journal} {\prl}\ }\textbf {\bibinfo
  {volume} {137}},\ \bibinfo {eid} {021404} (\bibinfo {year} {2026})},\ \Eprint
  {https://arxiv.org/abs/2601.07908} {arXiv:2601.07908 [gr-qc]} \BibitemShut
  {NoStop}%
\bibitem [{\citenamefont {{Tong}}\ \emph {et~al.}(2026)\citenamefont {{Tong}}
  \emph {et~al.}}]{2026Natur.652..874T}%
  \BibitemOpen
  \bibfield  {author} {\bibinfo {author} {\bibfnamefont {H.}~\bibnamefont
  {{Tong}}} \emph {et~al.},\ }\href
  {https://doi.org/10.1038/s41586-026-10359-0} {\bibfield  {journal} {\bibinfo
  {journal} {\nat}\ }\textbf {\bibinfo {volume} {652}},\ \bibinfo {pages} {874}
  (\bibinfo {year} {2026})},\ \Eprint {https://arxiv.org/abs/2509.04151}
  {arXiv:2509.04151 [astro-ph.HE]} \BibitemShut {NoStop}%
\bibitem [{\citenamefont {{Rinaldi}}\ \emph {et~al.}(2026)\citenamefont
  {{Rinaldi}}, \citenamefont {{Wong}}, \citenamefont {{Vaccaro}}, \citenamefont
  {{Mislimi}}, \citenamefont {{Hannuksela}}, \citenamefont {{Leong}},\ and\
  \citenamefont {{Mapelli}}}]{2026arXiv260623305R}%
  \BibitemOpen
  \bibfield  {author} {\bibinfo {author} {\bibfnamefont {S.}~\bibnamefont
  {{Rinaldi}}}, \bibinfo {author} {\bibfnamefont {C.}~\bibnamefont {{Wong}}},
  \bibinfo {author} {\bibfnamefont {M.~P.}\ \bibnamefont {{Vaccaro}}}, \bibinfo
  {author} {\bibfnamefont {E.}~\bibnamefont {{Mislimi}}}, \bibinfo {author}
  {\bibfnamefont {O.~A.}\ \bibnamefont {{Hannuksela}}}, \bibinfo {author}
  {\bibfnamefont {S.~H.~W.}\ \bibnamefont {{Leong}}},\ and\ \bibinfo {author}
  {\bibfnamefont {M.}~\bibnamefont {{Mapelli}}},\ }\href
  {https://doi.org/10.48550/arXiv.2606.23305} {\bibfield  {journal} {\bibinfo
  {journal} {arXiv e-prints}\ ,\ \bibinfo {eid} {arXiv:2606.23305}} (\bibinfo
  {year} {2026})},\ \Eprint {https://arxiv.org/abs/2606.23305}
  {arXiv:2606.23305 [astro-ph.HE]} \BibitemShut {NoStop}%
\bibitem [{\citenamefont {{Banagiri}}\ \emph {et~al.}(2026)\citenamefont
  {{Banagiri}}, \citenamefont {{Thrane}},\ and\ \citenamefont
  {{Lasky}}}]{2026PhRvL.137b1403B}%
  \BibitemOpen
  \bibfield  {author} {\bibinfo {author} {\bibfnamefont {S.}~\bibnamefont
  {{Banagiri}}}, \bibinfo {author} {\bibfnamefont {E.}~\bibnamefont
  {{Thrane}}},\ and\ \bibinfo {author} {\bibfnamefont {P.~D.}\ \bibnamefont
  {{Lasky}}},\ }\href {https://doi.org/10.1103/blyb-lqv6} {\bibfield  {journal}
  {\bibinfo  {journal} {\prl}\ }\textbf {\bibinfo {volume} {137}},\ \bibinfo
  {eid} {021403} (\bibinfo {year} {2026})},\ \Eprint
  {https://arxiv.org/abs/2509.15646} {arXiv:2509.15646 [astro-ph.HE]}
  \BibitemShut {NoStop}%
\bibitem [{\citenamefont {{Ray}}\ \emph {et~al.}(2026)\citenamefont {{Ray}},
  \citenamefont {{Mukherjee}}, \citenamefont {{Zevin}},\ and\ \citenamefont
  {{Kalogera}}}]{2026ApJ..1005L..55R}%
  \BibitemOpen
  \bibfield  {author} {\bibinfo {author} {\bibfnamefont {A.}~\bibnamefont
  {{Ray}}}, \bibinfo {author} {\bibfnamefont {S.}~\bibnamefont {{Mukherjee}}},
  \bibinfo {author} {\bibfnamefont {M.}~\bibnamefont {{Zevin}}},\ and\ \bibinfo
  {author} {\bibfnamefont {V.}~\bibnamefont {{Kalogera}}},\ }\href
  {https://doi.org/10.3847/2041-8213/ae80cb} {\bibfield  {journal} {\bibinfo
  {journal} {\apjl}\ }\textbf {\bibinfo {volume} {1005}},\ \bibinfo {eid} {L55}
  (\bibinfo {year} {2026})},\ \Eprint {https://arxiv.org/abs/2603.17987}
  {arXiv:2603.17987 [astro-ph.HE]} \BibitemShut {NoStop}%
\bibitem [{\citenamefont {{Adamcewicz}}\ \emph {et~al.}(2025)\citenamefont
  {{Adamcewicz}}, \citenamefont {{Guttman}}, \citenamefont {{Lasky}},\ and\
  \citenamefont {{Thrane}}}]{2025ApJ...994..261A}%
  \BibitemOpen
  \bibfield  {author} {\bibinfo {author} {\bibfnamefont {C.}~\bibnamefont
  {{Adamcewicz}}}, \bibinfo {author} {\bibfnamefont {N.}~\bibnamefont
  {{Guttman}}}, \bibinfo {author} {\bibfnamefont {P.~D.}\ \bibnamefont
  {{Lasky}}},\ and\ \bibinfo {author} {\bibfnamefont {E.}~\bibnamefont
  {{Thrane}}},\ }\href {https://doi.org/10.3847/1538-4357/ae1370} {\bibfield
  {journal} {\bibinfo  {journal} {\apj}\ }\textbf {\bibinfo {volume} {994}},\
  \bibinfo {eid} {261} (\bibinfo {year} {2025})},\ \Eprint
  {https://arxiv.org/abs/2509.04706} {arXiv:2509.04706 [astro-ph.HE]}
  \BibitemShut {NoStop}%
\bibitem [{\citenamefont {{Bartos}}\ and\ \citenamefont
  {{Haiman}}(2026)}]{2026arXiv260509351B}%
  \BibitemOpen
  \bibfield  {author} {\bibinfo {author} {\bibfnamefont {I.}~\bibnamefont
  {{Bartos}}}\ and\ \bibinfo {author} {\bibfnamefont {Z.}~\bibnamefont
  {{Haiman}}},\ }\href {https://doi.org/10.48550/arXiv.2605.09351} {\bibfield
  {journal} {\bibinfo  {journal} {arXiv e-prints}\ ,\ \bibinfo {eid}
  {arXiv:2605.09351}} (\bibinfo {year} {2026})},\ \Eprint
  {https://arxiv.org/abs/2605.09351} {arXiv:2605.09351 [astro-ph.HE]}
  \BibitemShut {NoStop}%
\bibitem [{\citenamefont {{Hussain}}\ \emph {et~al.}(2026)\citenamefont
  {{Hussain}}, \citenamefont {{Isi}},\ and\ \citenamefont
  {{Zimmerman}}}]{2026arXiv260524281H}%
  \BibitemOpen
  \bibfield  {author} {\bibinfo {author} {\bibfnamefont {A.}~\bibnamefont
  {{Hussain}}}, \bibinfo {author} {\bibfnamefont {M.}~\bibnamefont {{Isi}}},\
  and\ \bibinfo {author} {\bibfnamefont {A.}~\bibnamefont {{Zimmerman}}},\
  }\href {https://doi.org/10.48550/arXiv.2605.24281} {\bibfield  {journal}
  {\bibinfo  {journal} {arXiv e-prints}\ ,\ \bibinfo {eid} {arXiv:2605.24281}}
  (\bibinfo {year} {2026})},\ \Eprint {https://arxiv.org/abs/2605.24281}
  {arXiv:2605.24281 [astro-ph.HE]} \BibitemShut {NoStop}%
\bibitem [{\citenamefont {{Roupas}}(2025)}]{2025A&A...702A.208R}%
  \BibitemOpen
  \bibfield  {author} {\bibinfo {author} {\bibfnamefont {Z.}~\bibnamefont
  {{Roupas}}},\ }\href {https://doi.org/10.1051/0004-6361/202556434} {\bibfield
   {journal} {\bibinfo  {journal} {\aap}\ }\textbf {\bibinfo {volume} {702}},\
  \bibinfo {eid} {A208} (\bibinfo {year} {2025})},\ \Eprint
  {https://arxiv.org/abs/2509.08448} {arXiv:2509.08448 [astro-ph.GA]}
  \BibitemShut {NoStop}%
\bibitem [{\citenamefont {{Roupas}}(2026{\natexlab{a}})}]{2026A&A...709A.5R}%
  \BibitemOpen
  \bibfield  {author} {\bibinfo {author} {\bibfnamefont {Z.}~\bibnamefont
  {{Roupas}}},\ }\href {https://doi.org/10.1051/0004-6361/202558435} {\bibfield
   {journal} {\bibinfo  {journal} {\aap}\ }\textbf {\bibinfo {volume} {709}},\
  \bibinfo {eid} {A5} (\bibinfo {year} {2026}{\natexlab{a}})},\ \Eprint
  {https://arxiv.org/abs/2603.18857} {arXiv:2603.18857 [astro-ph.GA]}
  \BibitemShut {NoStop}%
\bibitem [{\citenamefont {{Roupas}}(2026{\natexlab{b}})}]{2026arXiv260712465R}%
  \BibitemOpen
  \bibfield  {author} {\bibinfo {author} {\bibfnamefont {Z.}~\bibnamefont
  {{Roupas}}},\ }\href {https://doi.org/10.48550/arXiv.2607.12465} {\bibfield
  {journal} {\bibinfo  {journal} {arXiv e-prints}\ ,\ \bibinfo {eid}
  {arXiv:2607.12465}} (\bibinfo {year} {2026}{\natexlab{b}})},\ \Eprint
  {https://arxiv.org/abs/2607.12465} {arXiv:2607.12465 [astro-ph.GA]}
  \BibitemShut {NoStop}%
\bibitem [{\citenamefont {{Adamo}}\ \emph {et~al.}(2024)\citenamefont {{Adamo}}
  \emph {et~al.}}]{2024Natur.632..513A}%
  \BibitemOpen
  \bibfield  {author} {\bibinfo {author} {\bibfnamefont {A.}~\bibnamefont
  {{Adamo}}} \emph {et~al.},\ }\href
  {https://doi.org/10.1038/s41586-024-07703-7} {\bibfield  {journal} {\bibinfo
  {journal} {\nat}\ }\textbf {\bibinfo {volume} {632}},\ \bibinfo {pages} {513}
  (\bibinfo {year} {2024})},\ \Eprint {https://arxiv.org/abs/2401.03224}
  {arXiv:2401.03224 [astro-ph.GA]} \BibitemShut {NoStop}%
\bibitem [{\citenamefont {{Roupas}}(2026{\natexlab{c}})}]{2026arXiv260717869R}%
  \BibitemOpen
  \bibfield  {author} {\bibinfo {author} {\bibfnamefont {Z.}~\bibnamefont
  {{Roupas}}},\ }\href {https://doi.org/10.48550/arXiv.2607.17869} {\bibfield
  {journal} {\bibinfo  {journal} {arXiv e-prints}\ ,\ \bibinfo {eid}
  {arXiv:2607.17869}} (\bibinfo {year} {2026}{\natexlab{c}})},\ \Eprint
  {https://arxiv.org/abs/2607.17869} {arXiv:2607.17869 [astro-ph.HE]}
  \BibitemShut {NoStop}%
\bibitem [{\citenamefont {{Abac}}\ \emph {et~al.}(2026)\citenamefont {{Abac}}
  \emph {et~al.}}]{2026ApJ..1005L..51A}%
  \BibitemOpen
  \bibfield  {author} {\bibinfo {author} {\bibfnamefont {A.~G.}\ \bibnamefont
  {{Abac}}} \emph {et~al.},\ }\href {https://doi.org/10.3847/2041-8213/ae771e}
  {\bibfield  {journal} {\bibinfo  {journal} {\apjl}\ }\textbf {\bibinfo
  {volume} {1005}},\ \bibinfo {eid} {L51} (\bibinfo {year} {2026})},\ \Eprint
  {https://arxiv.org/abs/2508.18083} {arXiv:2508.18083 [astro-ph.HE]}
  \BibitemShut {NoStop}%
\bibitem [{\citenamefont {{Berti}}\ and\ \citenamefont
  {{Volonteri}}(2008)}]{2008ApJ...684..822B}%
  \BibitemOpen
  \bibfield  {author} {\bibinfo {author} {\bibfnamefont {E.}~\bibnamefont
  {{Berti}}}\ and\ \bibinfo {author} {\bibfnamefont {M.}~\bibnamefont
  {{Volonteri}}},\ }\href {https://doi.org/10.1086/590379} {\bibfield
  {journal} {\bibinfo  {journal} {\apj}\ }\textbf {\bibinfo {volume} {684}},\
  \bibinfo {pages} {822} (\bibinfo {year} {2008})},\ \Eprint
  {https://arxiv.org/abs/0802.0025} {arXiv:0802.0025 [astro-ph]} \BibitemShut
  {NoStop}%
\bibitem [{\citenamefont {{Fishbach}}\ \emph {et~al.}(2017)\citenamefont
  {{Fishbach}}, \citenamefont {{Holz}},\ and\ \citenamefont
  {{Farr}}}]{2017ApJ...840L..24F}%
  \BibitemOpen
  \bibfield  {author} {\bibinfo {author} {\bibfnamefont {M.}~\bibnamefont
  {{Fishbach}}}, \bibinfo {author} {\bibfnamefont {D.~E.}\ \bibnamefont
  {{Holz}}},\ and\ \bibinfo {author} {\bibfnamefont {B.}~\bibnamefont
  {{Farr}}},\ }\href {https://doi.org/10.3847/2041-8213/aa7045} {\bibfield
  {journal} {\bibinfo  {journal} {\apjl}\ }\textbf {\bibinfo {volume} {840}},\
  \bibinfo {eid} {L24} (\bibinfo {year} {2017})},\ \Eprint
  {https://arxiv.org/abs/1703.06869} {arXiv:1703.06869 [astro-ph.HE]}
  \BibitemShut {NoStop}%
\bibitem [{\citenamefont {{Ma}}\ \emph {et~al.}(2020)\citenamefont {{Ma}} \emph
  {et~al.}}]{2020MNRAS.493.4315M}%
  \BibitemOpen
  \bibfield  {author} {\bibinfo {author} {\bibfnamefont {X.}~\bibnamefont
  {{Ma}}} \emph {et~al.},\ }\href {https://doi.org/10.1093/mnras/staa527}
  {\bibfield  {journal} {\bibinfo  {journal} {\mnras}\ }\textbf {\bibinfo
  {volume} {493}},\ \bibinfo {pages} {4315} (\bibinfo {year} {2020})},\ \Eprint
  {https://arxiv.org/abs/1906.11261} {arXiv:1906.11261 [astro-ph.GA]}
  \BibitemShut {NoStop}%
\bibitem [{\citenamefont {{Pascale}}\ \emph {et~al.}(2025)\citenamefont
  {{Pascale}} \emph {et~al.}}]{2025A&A...699A..31P}%
  \BibitemOpen
  \bibfield  {author} {\bibinfo {author} {\bibfnamefont {R.}~\bibnamefont
  {{Pascale}}} \emph {et~al.},\ }\href
  {https://doi.org/10.1051/0004-6361/202453252} {\bibfield  {journal} {\bibinfo
   {journal} {\aap}\ }\textbf {\bibinfo {volume} {699}},\ \bibinfo {eid} {A31}
  (\bibinfo {year} {2025})},\ \Eprint {https://arxiv.org/abs/2505.06346}
  {arXiv:2505.06346 [astro-ph.GA]} \BibitemShut {NoStop}%
\bibitem [{\citenamefont {{Adamo}}\ \emph {et~al.}(2020)\citenamefont {{Adamo}}
  \emph {et~al.}}]{2020MNRAS.499.3267A}%
  \BibitemOpen
  \bibfield  {author} {\bibinfo {author} {\bibfnamefont {A.}~\bibnamefont
  {{Adamo}}} \emph {et~al.},\ }\href {https://doi.org/10.1093/mnras/staa2380}
  {\bibfield  {journal} {\bibinfo  {journal} {\mnras}\ }\textbf {\bibinfo
  {volume} {499}},\ \bibinfo {pages} {3267} (\bibinfo {year} {2020})},\ \Eprint
  {https://arxiv.org/abs/2008.12794} {arXiv:2008.12794 [astro-ph.GA]}
  \BibitemShut {NoStop}%
\bibitem [{\citenamefont {{Linden}}\ \emph {et~al.}(2021)\citenamefont
  {{Linden}} \emph {et~al.}}]{2021ApJ...923..278L}%
  \BibitemOpen
  \bibfield  {author} {\bibinfo {author} {\bibfnamefont {S.~T.}\ \bibnamefont
  {{Linden}}} \emph {et~al.},\ }\href
  {https://doi.org/10.3847/1538-4357/ac2892} {\bibfield  {journal} {\bibinfo
  {journal} {\apj}\ }\textbf {\bibinfo {volume} {923}},\ \bibinfo {eid} {278}
  (\bibinfo {year} {2021})},\ \Eprint {https://arxiv.org/abs/2110.03638}
  {arXiv:2110.03638 [astro-ph.GA]} \BibitemShut {NoStop}%
\bibitem [{\citenamefont {{Fuller}}\ and\ \citenamefont
  {{Ma}}(2019)}]{2019ApJ...881L...1F}%
  \BibitemOpen
  \bibfield  {author} {\bibinfo {author} {\bibfnamefont {J.}~\bibnamefont
  {{Fuller}}}\ and\ \bibinfo {author} {\bibfnamefont {L.}~\bibnamefont
  {{Ma}}},\ }\href {https://doi.org/10.3847/2041-8213/ab339b} {\bibfield
  {journal} {\bibinfo  {journal} {\apjl}\ }\textbf {\bibinfo {volume} {881}},\
  \bibinfo {eid} {L1} (\bibinfo {year} {2019})},\ \Eprint
  {https://arxiv.org/abs/1907.03714} {arXiv:1907.03714 [astro-ph.SR]}
  \BibitemShut {NoStop}%
\bibitem [{\citenamefont {Fuller}\ \emph {et~al.}(2019)\citenamefont {Fuller},
  \citenamefont {Piro},\ and\ \citenamefont {Jermyn}}]{Fuller_2019}%
  \BibitemOpen
  \bibfield  {author} {\bibinfo {author} {\bibfnamefont {J.}~\bibnamefont
  {Fuller}}, \bibinfo {author} {\bibfnamefont {A.~L.}\ \bibnamefont {Piro}},\
  and\ \bibinfo {author} {\bibfnamefont {A.~S.}\ \bibnamefont {Jermyn}},\
  }\href {https://doi.org/10.1093/mnras/stz514} {\bibfield  {journal} {\bibinfo
   {journal} {Monthly Notices of the Royal Astronomical Society}\ }\textbf
  {\bibinfo {volume} {485}},\ \bibinfo {pages} {3661–3680} (\bibinfo {year}
  {2019})}\BibitemShut {NoStop}%
\bibitem [{\citenamefont {{Belczynski}}\ \emph {et~al.}(2020)\citenamefont
  {{Belczynski}} \emph {et~al.}}]{2020A&A...636A.104B}%
  \BibitemOpen
  \bibfield  {author} {\bibinfo {author} {\bibfnamefont {K.}~\bibnamefont
  {{Belczynski}}} \emph {et~al.},\ }\href
  {https://doi.org/10.1051/0004-6361/201936528} {\bibfield  {journal} {\bibinfo
   {journal} {\aap}\ }\textbf {\bibinfo {volume} {636}},\ \bibinfo {eid} {A104}
  (\bibinfo {year} {2020})},\ \Eprint {https://arxiv.org/abs/1706.07053}
  {arXiv:1706.07053 [astro-ph.HE]} \BibitemShut {NoStop}%
\bibitem [{\citenamefont {Spruit}(2002)}]{Spruit_2002}%
  \BibitemOpen
  \bibfield  {author} {\bibinfo {author} {\bibfnamefont {H.~C.}\ \bibnamefont
  {Spruit}},\ }\href {https://doi.org/10.1051/0004-6361:20011465} {\bibfield
  {journal} {\bibinfo  {journal} {\aap}\ }\textbf {\bibinfo {volume} {381}},\
  \bibinfo {pages} {923–932} (\bibinfo {year} {2002})}\BibitemShut {NoStop}%
\bibitem [{\citenamefont {Heger}\ \emph {et~al.}(2005)\citenamefont {Heger},
  \citenamefont {Woosley},\ and\ \citenamefont {Spruit}}]{Heger_2005}%
  \BibitemOpen
  \bibfield  {author} {\bibinfo {author} {\bibfnamefont {A.}~\bibnamefont
  {Heger}}, \bibinfo {author} {\bibfnamefont {S.~E.}\ \bibnamefont {Woosley}},\
  and\ \bibinfo {author} {\bibfnamefont {H.~C.}\ \bibnamefont {Spruit}},\
  }\href {https://doi.org/10.1086/429868} {\bibfield  {journal} {\bibinfo
  {journal} {\apj}\ }\textbf {\bibinfo {volume} {626}},\ \bibinfo {pages}
  {350–363} (\bibinfo {year} {2005})}\BibitemShut {NoStop}%
\bibitem [{\citenamefont {Qin}\ \emph {et~al.}(2018)\citenamefont {Qin} \emph
  {et~al.}}]{Qin_2018}%
  \BibitemOpen
  \bibfield  {author} {\bibinfo {author} {\bibfnamefont {Y.}~\bibnamefont
  {Qin}} \emph {et~al.},\ }\href {https://doi.org/10.1051/0004-6361/201832839}
  {\bibfield  {journal} {\bibinfo  {journal} {\aap}\ }\textbf {\bibinfo
  {volume} {616}},\ \bibinfo {pages} {A28} (\bibinfo {year}
  {2018})}\BibitemShut {NoStop}%
\end{thebibliography}%
	\bibliographystyle{apsrev4-2}	
		
\appendix		

\section{Star-cluster modelling}\label{app:clusters}
Direct observational evidence of proto-globular clusters (GCs) with mass $\sim 10^6\,{\rm M}_{\odot}$, and size $\sim 1\,{\rm pc}$, at high redshift $\sim 10$---indicating  subsolar metallicity---comes from the James Webb Space Telescope in the Cosmic Gems arc galaxy \citep{2024Natur.632..513A}. 
A spin-mass correlation of BHs is produced for sufficiently massive and dense clusters, comparable to the Cosmic Gems arc clusters, which can produce BH masses $\gtrsim 10^2\,{\rm M}_{\odot}$ \citep{2026A&A...709A.5R}. 
We wish to identify candidate GW events whose BH components could have originated in clusters capable of generating BH masses comparable at least to the most massive BH component observed to date.
We get realistic cluster densities for subsolar metallicities, which imply weaker stellar winds and therefore longer depletion timescales. 
Therefore, we use simulation data from Ref.~\citep{2026A&A...709A.5R} for clusters able to generate masses of at least $150\,{\rm M}_{\odot}$ and up to the maximum attainable with this mechanism, $\sim 10^{3}\,{\rm M}_{\odot}$. We consider star clusters with masses $M_{\star} = (0.35-1) \cdot 10^{6}\,{\rm M}_\odot$, size $r_{c,\star} = 0.5-1.2\,{\rm pc}$, two metallicity values $Z = \{0.01,0.1\}\,Z_{\odot}$, and star formation efficiency $0.35$, consistent with a subsolar metallicity. 

\begin{figure}
	\centering
	\centering
	\includegraphics[width=0.95\columnwidth]{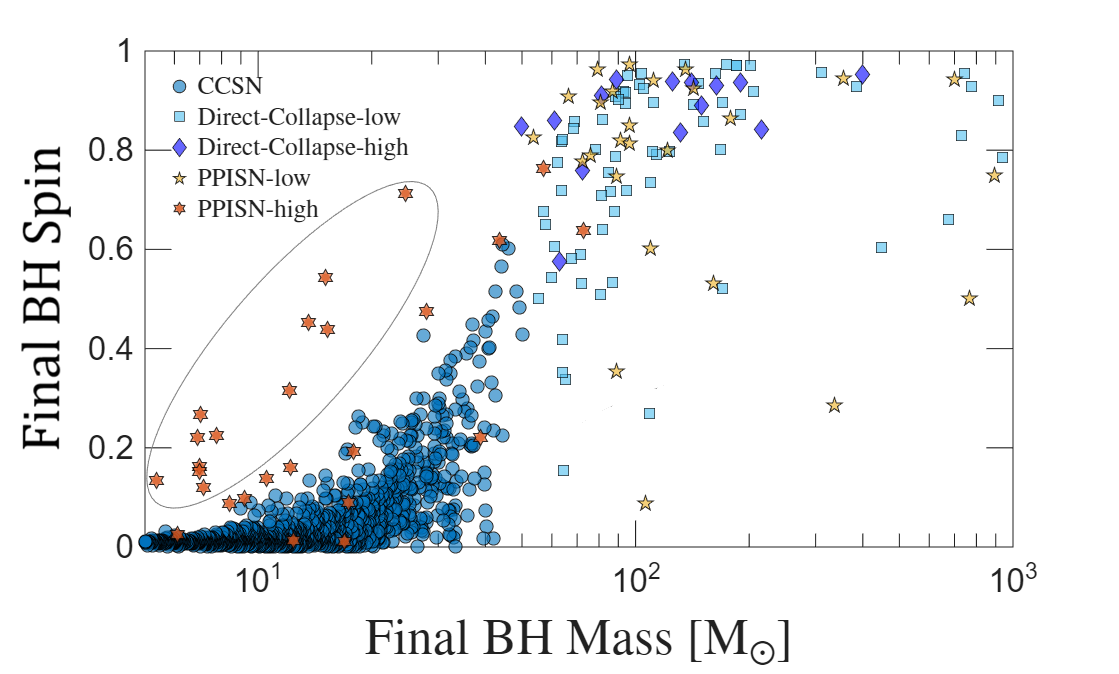}
	\caption{Final BH spin with respect to final BH mass of all BHs for one representative simulation run of our typical cluster with $M_{\star} = 10^6\,{\rm M}_\odot$, $r_{c,\star} = 1 \,{\rm pc}$. We have circled low-mass/intermediate-spin outliers. The legend labels provide the BH origin; CCSN for a core-collapse-supernova remnant; Direct-Collapse-low(high) for a BH generated by direct collapse of the core of a low(high) mass star; PPISN-low(high) for a BH remnant from a pulsational-pair-instability supernova of low(high) mass. From the simulations of \citep{2026A&A...709A.5R}.}
	\label{fig:a_m_fin_data}
\end{figure}

Cosmological simulations focused on investigating the formation of proto-GCs at high $z$ reveal a star-cluster mass function $dN/dM_{\rm cl} \propto M_{\rm cl}^{-\alpha}$, $\alpha = 2$,  \citep{2020MNRAS.493.4315M,2025A&A...699A..31P}
along a vast range of proto-GC masses spanning the range of our inspected values. Local observations for young starbursts are consistent with an exponent in the range $\alpha \sim 1.5-2.0$ (e.g. \citep{2020MNRAS.499.3267A,2021ApJ...923..278L}).
We find that our analysis is not sensitive to an exponent in this range. We adopt the value $\alpha=2$ as the more conservative choice for our purposes---giving a $15\%$ lower population evidence than the value $1.5$---supported also by simulations of proto-GCs, the class of clusters we consider. We adopt equal weights for the several cluster sizes and the two metallicity values, given the uncertainty of those values. Nevertheless, the form of the spin-mass correlation is not sensitive to those parameters as long as they refer to clusters which can produce BHs with masses up to $\sim 150\,{\rm M}_{\odot}$, or above. In this case, the median spin can be consistently fitted with a saturated exponential law of the form Eq.~(\ref{eq:med_fit}) \citep{2026A&A...709A.5R}.

This general form refers to the adopted natal BH spin at formation of $\chi_{\rm ini}=0.01$ used in our cluster simulations, consistent with the recent models of efficient angular momentum transport during core contraction of the massive star progenitor (revised Tayler instability) \citep{2019ApJ...881L...1F,Fuller_2019,2020A&A...636A.104B}.
Even in the case of natal spin $0.1$, consistent with the classic Tayler-Spruit dynamo \citep{Spruit_2002,Heger_2005,Qin_2018}, the general form (\ref{eq:med_fit}) persists with $\chi$ only shifted to values $\sim 0.1$ or lower for $m\lesssim 20\,{\rm M}_{\odot}$. The maximum spin, the transition region, and the exponent remain nearly identical to our adopted $\chi_{\rm ini} = 0.01$ case. In fact the case $0.1$ matches the GW data at low masses better than our adopted value. Therefore, our choice can be regarded as a conservative assumption for our purposes.

In Figure~\ref{fig:a_m_fin_data} are depicted the final spins and masses, after gas depletion, of all BHs in one representative star cluster of our simulations. 
The more massive pulsational-pair-instability supernova leave lower-mass BHs, which nevertheless are born very early originating in the most massive stars. They have therefore a significant time-window available to accrete angular momentum from the gas. They remain to low-to-medium masses because they have initially very low mass ($\sim 5\,{\rm M}_{\odot}$), and also tend to move outside the deeper part of the cluster core, being more sensitive to random gravitational fluctuations than the initially more massive BHs. Thus, they form a population of low-mass/intermediate-spin outliers encircled in the Figure.

\onecolumngrid

\section{Complete event table}

	% longtable's own caption style differs from the class's; this restores it.
	\makeatletter
	\def\LT@makecaption#1#2#3{%
		\LT@mcol\LT@cols c{\parbox{\LTcapwidth}{\@makecaption{#2}{#3}}}}
	\makeatother
	\squeezetable
	\setlength{\LTcapwidth}{0.95\textwidth}
	\addtolength{\LTcapwidth}{-\columnsep}
	\begin{longtable}{lrrrrrrrrrrrl}
		\caption{%
			Complete GWTC-5.0 binary-black-hole sample ($N=259$) analyzed in this
			work, in chronological order. 
			$m_1,m_2$: medians of source-frame component masses $[{\rm M_\odot}]$;
			$\chi_1,\chi_2$: medians of component spin magnitudes;
			$\chi_{\rm eff}$: median effective inspiral spin;
			$\Delta_{90}\chi_i$: width of the $90\%$ credible interval of $\chi_i$;
			$\ln B^{\rm Def}$, $\ln B^{\rm 2br}$, $\ln B^{\rm Merg}$: log Bayes
			factor of accretion against the \textsc{Default-bbh} model, the
			\textsc{2br} spin sector, and against the hierarchical-merger channel
			\textsc{Merg}, with their sampling uncertainties;
			$P^{\rm Def}_{\rm acc}$: accretion membership probability;
			Class: `acc' for $\ln B^{\rm Def}>1$,  `acc weak' for
			$0.5<\ln B^{\rm Def}\le1$, `no pref' for
			$|\ln B^{\rm Def}|\le0.5$, and `pop' for
			$\ln B^{\rm Def}<-0.5$.}
		\label{tab:full_catalogue}
		\\
		\hline\hline
	Event & $m_1$ & $m_2$ & $\chi_1$ & $\chi_2$ & $\chi_{\rm eff}$ &
		$\Delta_{90}\chi_1$ & $\Delta_{90}\chi_2$ &
		$\ln B^{\rm Def}$ & $\ln B^{\rm 2br}$ & $\ln B^{\rm Merg}$ &
		$P^{\rm Def}_{\rm acc}$ & Class
		\\
		\midrule
		\endfirsthead

		\multicolumn{13}{l}{\small\itshape Table~\ref{tab:full_catalogue} (continued)}
		\\
		\hline\hline
	Event & $m_1$ & $m_2$ & $\chi_1$ & $\chi_2$ & $\chi_{\rm eff}$ &
		$\Delta_{90}\chi_1$ & $\Delta_{90}\chi_2$ &
		$\ln B^{\rm Def}$ & $\ln B^{\rm 2br}$ & $\ln B^{\rm Merg}$ &
		$P^{\rm Def}_{\rm acc}$ & Class
		\\
		\midrule
		\endhead

		\hline\hline
		\endlastfoot

		GW150914\_095045 & $34.60$ & $30.02$ & $0.454$ & $0.420$ & $-0.038$ & $0.841$ & $0.867$ & $+0.081\pm0.038$ & $+0.064\pm0.039$ & $+0.096\pm0.038$ & $0.3998$ & \mbox{no pref} \\
		GW151012\_095443 & $24.78$ & $13.59$ & $0.385$ & $0.460$ & $0.119$ & $0.836$ & $0.876$ & $+0.151\pm0.088$ & $+0.130\pm0.088$ & $+0.171\pm0.088$ & $0.4162$ & \mbox{no pref} \\
		GW151226\_033853 & $14.17$ & $7.46$ & $0.607$ & $0.453$ & $0.202$ & $0.807$ & $0.875$ & $-4.460\pm0.029$ & $-4.425\pm0.029$ & $-4.497\pm0.029$ & $0.0075$ & \mbox{pop} \\
		GW170104\_101158 & $28.71$ & $20.79$ & $0.345$ & $0.381$ & $-0.042$ & $0.808$ & $0.867$ & $+0.284\pm0.021$ & $+0.251\pm0.021$ & $+0.320\pm0.021$ & $0.4475$ & \mbox{no pref} \\
		GW170608\_020116 & $10.64$ & $7.82$ & $0.294$ & $0.313$ & $0.045$ & $0.706$ & $0.802$ & $+0.331\pm0.105$ & $+0.282\pm0.105$ & $+0.356\pm0.105$ & $0.4586$ & \mbox{no pref} \\
		GW170729\_185629 & $54.68$ & $30.20$ & $0.603$ & $0.480$ & $0.289$ & $0.885$ & $0.897$ & $+0.297\pm0.023$ & $+0.205\pm0.021$ & $+0.274\pm0.022$ & $0.4505$ & \mbox{no pref} \\
		GW170809\_082821 & $34.14$ & $24.21$ & $0.351$ & $0.401$ & $0.074$ & $0.833$ & $0.869$ & $+0.118\pm0.015$ & $+0.089\pm0.015$ & $+0.152\pm0.015$ & $0.4085$ & \mbox{no pref} \\
		GW170814\_103043 & $30.92$ & $24.93$ & $0.412$ & $0.419$ & $0.082$ & $0.857$ & $0.875$ & $+0.170\pm0.019$ & $+0.150\pm0.019$ & $+0.191\pm0.018$ & $0.4206$ & \mbox{no pref} \\
		GW170818\_022509 & $34.78$ & $27.60$ & $0.522$ & $0.475$ & $-0.057$ & $0.882$ & $0.888$ & $-0.005\pm0.016$ & $-0.003\pm0.016$ & $-0.012\pm0.016$ & $0.3803$ & \mbox{no pref} \\
		GW170823\_131358 & $38.26$ & $28.98$ & $0.435$ & $0.430$ & $0.052$ & $0.874$ & $0.878$ & $+0.062\pm0.022$ & $+0.054\pm0.022$ & $+0.080\pm0.022$ & $0.3955$ & \mbox{no pref} \\
		GW190408\_181802 & $24.81$ & $18.53$ & $0.306$ & $0.371$ & $-0.026$ & $0.784$ & $0.857$ & $+0.495\pm0.046$ & $+0.457\pm0.046$ & $+0.537\pm0.046$ & $0.4978$ & \mbox{no pref} \\
		GW190412\_053044 & $27.75$ & $9.02$ & $0.315$ & $0.406$ & $0.214$ & $0.411$ & $0.869$ & $-0.039\pm0.065$ & $-0.081\pm0.065$ & $+0.009\pm0.065$ & $0.3725$ & \mbox{no pref} \\
		GW190413\_052954 & $33.70$ & $24.21$ & $0.416$ & $0.453$ & $-0.035$ & $0.878$ & $0.882$ & $+0.047\pm0.020$ & $+0.032\pm0.020$ & $+0.063\pm0.020$ & $0.3920$ & \mbox{no pref} \\
		GW190413\_134308 & $51.15$ & $30.53$ & $0.597$ & $0.490$ & $-0.007$ & $0.895$ & $0.890$ & $+0.282\pm0.020$ & $+0.111\pm0.019$ & $+0.270\pm0.020$ & $0.4471$ & \mbox{no pref} \\
		GW190421\_213856 & $42.02$ & $32.03$ & $0.432$ & $0.459$ & $-0.095$ & $0.880$ & $0.885$ & $-0.044\pm0.014$ & $-0.020\pm0.014$ & $-0.030\pm0.014$ & $0.3715$ & \mbox{no pref} \\
		GW190503\_185404 & $41.30$ & $28.31$ & $0.429$ & $0.441$ & $-0.045$ & $0.864$ & $0.887$ & $-0.011\pm0.017$ & $+0.007\pm0.017$ & $+0.005\pm0.017$ & $0.3789$ & \mbox{no pref} \\
		GW190512\_180714 & $23.15$ & $12.53$ & $0.203$ & $0.405$ & $0.024$ & $0.675$ & $0.871$ & $+0.732\pm0.048$ & $+0.684\pm0.048$ & $+0.785\pm0.047$ & $0.5544$ & \mbox{acc weak} \\
		GW190513\_205428 & $35.96$ & $18.32$ & $0.417$ & $0.455$ & $0.163$ & $0.839$ & $0.884$ & $+0.142\pm0.038$ & $+0.125\pm0.038$ & $+0.154\pm0.037$ & $0.4141$ & \mbox{no pref} \\
		GW190517\_055101 & $39.24$ & $24.02$ & $0.896$ & $0.617$ & $0.491$ & $0.394$ & $0.885$ & $-0.079\pm0.078$ & $-0.665\pm0.078$ & $-0.287\pm0.078$ & $0.3636$ & \mbox{no pref} \\
		GW190519\_153544 & $65.10$ & $40.83$ & $0.609$ & $0.588$ & $0.331$ & $0.795$ & $0.887$ & $+0.791\pm0.021$ & $+0.231\pm0.019$ & $+0.697\pm0.021$ & $0.5683$ & \mbox{acc weak} \\
		GW190521\_030229 & $98.40$ & $57.25$ & $0.710$ & $0.534$ & $-0.140$ & $0.890$ & $0.898$ & $+0.892\pm0.017$ & $+0.526\pm0.015$ & $+0.867\pm0.016$ & $0.5919$ & \mbox{acc weak} \\
		GW190521\_074359 & $43.39$ & $33.43$ & $0.326$ & $0.425$ & $0.096$ & $0.774$ & $0.855$ & $-0.179\pm0.023$ & $-0.093\pm0.023$ & $-0.147\pm0.023$ & $0.3415$ & \mbox{no pref} \\
		GW190527\_092055 & $35.60$ & $22.16$ & $0.371$ & $0.394$ & $0.096$ & $0.861$ & $0.876$ & $+0.082\pm0.028$ & $+0.063\pm0.028$ & $+0.111\pm0.028$ & $0.4001$ & \mbox{no pref} \\
		GW190602\_175927 & $71.91$ & $44.80$ & $0.464$ & $0.518$ & $0.116$ & $0.885$ & $0.891$ & $-0.026\pm0.015$ & $-0.014\pm0.014$ & $-0.029\pm0.015$ & $0.3755$ & \mbox{no pref} \\
		GW190620\_030421 & $58.02$ & $35.00$ & $0.696$ & $0.570$ & $0.340$ & $0.818$ & $0.895$ & $+0.967\pm0.031$ & $+0.327\pm0.028$ & $+0.863\pm0.030$ & $0.6092$ & \mbox{acc weak} \\
		GW190630\_185205 & $35.14$ & $23.97$ & $0.270$ & $0.427$ & $0.101$ & $0.687$ & $0.861$ & $+0.041\pm0.021$ & $+0.013\pm0.021$ & $+0.080\pm0.021$ & $0.3908$ & \mbox{no pref} \\
		GW190701\_203306 & $54.10$ & $40.50$ & $0.438$ & $0.452$ & $-0.082$ & $0.879$ & $0.890$ & $-0.163\pm0.015$ & $-0.024\pm0.014$ & $-0.150\pm0.015$ & $0.3450$ & \mbox{no pref} \\
		GW190706\_222641 & $73.97$ & $39.35$ & $0.637$ & $0.515$ & $0.282$ & $0.825$ & $0.892$ & $+0.635\pm0.031$ & $+0.259\pm0.027$ & $+0.582\pm0.030$ & $0.5314$ & \mbox{acc weak} \\
		GW190707\_093326 & $12.07$ & $7.93$ & $0.197$ & $0.334$ & $-0.037$ & $0.650$ & $0.846$ & $+1.153\pm0.090$ & $+1.098\pm0.090$ & $+1.194\pm0.090$ & $0.6508$ & \mbox{acc} \\
		GW190708\_232457 & $19.81$ & $11.61$ & $0.226$ & $0.335$ & $0.054$ & $0.677$ & $0.829$ & $+0.641\pm0.064$ & $+0.588\pm0.064$ & $+0.701\pm0.064$ & $0.5329$ & \mbox{acc weak} \\
		GW190719\_215514 & $36.78$ & $19.82$ & $0.602$ & $0.519$ & $0.251$ & $0.869$ & $0.890$ & $+0.208\pm0.045$ & $+0.101\pm0.045$ & $+0.179\pm0.045$ & $0.4295$ & \mbox{no pref} \\
		GW190720\_000836 & $14.21$ & $7.48$ & $0.357$ & $0.473$ & $0.193$ & $0.677$ & $0.872$ & $-3.310\pm0.159$ & $-3.320\pm0.159$ & $-3.289\pm0.159$ & $0.0233$ & \mbox{pop} \\
		GW190725\_174728 & $11.81$ & $6.29$ & $0.352$ & $0.494$ & $-0.039$ & $0.837$ & $0.887$ & $+0.138\pm0.098$ & $+0.119\pm0.099$ & $+0.158\pm0.098$ & $0.4132$ & \mbox{no pref} \\
		GW190727\_060333 & $38.90$ & $30.15$ & $0.502$ & $0.467$ & $0.094$ & $0.884$ & $0.883$ & $+0.057\pm0.024$ & $+0.041\pm0.024$ & $+0.057\pm0.024$ & $0.3944$ & \mbox{no pref} \\
		GW190728\_064510 & $12.45$ & $8.03$ & $0.333$ & $0.374$ & $0.133$ & $0.671$ & $0.845$ & $-3.261\pm0.093$ & $-3.296\pm0.093$ & $-3.232\pm0.093$ & $0.0245$ & \mbox{pop} \\
		GW190731\_140936 & $41.81$ & $29.02$ & $0.389$ & $0.458$ & $0.074$ & $0.879$ & $0.884$ & $-0.087\pm0.036$ & $-0.050\pm0.036$ & $-0.067\pm0.036$ & $0.3618$ & \mbox{no pref} \\
		GW190803\_022701 & $37.58$ & $27.69$ & $0.414$ & $0.440$ & $-0.015$ & $0.878$ & $0.883$ & $+0.018\pm0.019$ & $+0.010\pm0.019$ & $+0.038\pm0.018$ & $0.3853$ & \mbox{no pref} \\
		GW190805\_211137 & $46.17$ & $30.63$ & $0.755$ & $0.587$ & $0.368$ & $0.815$ & $0.891$ & $+0.554\pm0.048$ & $+0.158\pm0.045$ & $+0.471\pm0.047$ & $0.5119$ & \mbox{acc weak} \\
		GW190828\_063405 & $31.92$ & $25.81$ & $0.439$ & $0.406$ & $0.154$ & $0.865$ & $0.862$ & $+0.074\pm0.014$ & $+0.058\pm0.014$ & $+0.091\pm0.014$ & $0.3982$ & \mbox{no pref} \\
		GW190828\_065509 & $23.66$ & $10.44$ & $0.266$ & $0.437$ & $0.054$ & $0.686$ & $0.882$ & $+0.373\pm0.032$ & $+0.333\pm0.032$ & $+0.416\pm0.032$ & $0.4685$ & \mbox{no pref} \\
		GW190910\_112807 & $43.80$ & $34.15$ & $0.319$ & $0.365$ & $-0.003$ & $0.811$ & $0.860$ & $-0.227\pm0.015$ & $-0.117\pm0.015$ & $-0.185\pm0.015$ & $0.3312$ & \mbox{no pref} \\
		GW190915\_235702 & $32.57$ & $24.45$ & $0.551$ & $0.449$ & $-0.025$ & $0.876$ & $0.887$ & $+0.067\pm0.025$ & $+0.068\pm0.025$ & $+0.055\pm0.025$ & $0.3967$ & \mbox{no pref} \\
		GW190924\_021846 & $8.78$ & $5.11$ & $0.228$ & $0.328$ & $0.025$ & $0.664$ & $0.828$ & $+0.695\pm0.111$ & $+0.641\pm0.111$ & $+0.702\pm0.111$ & $0.5457$ & \mbox{acc weak} \\
		GW190925\_232845 & $20.84$ & $15.55$ & $0.355$ & $0.411$ & $0.086$ & $0.820$ & $0.872$ & $+0.247\pm0.050$ & $+0.217\pm0.050$ & $+0.279\pm0.050$ & $0.4386$ & \mbox{no pref} \\
		GW190929\_012149 & $66.29$ & $26.78$ & $0.338$ & $0.474$ & $-0.032$ & $0.846$ & $0.889$ & $-0.519\pm0.019$ & $-0.238\pm0.018$ & $-0.494\pm0.018$ & $0.2720$ & \mbox{pop} \\
		GW190930\_133541 & $14.19$ & $6.91$ & $0.391$ & $0.457$ & $0.193$ & $0.731$ & $0.874$ & $-1.608\pm0.252$ & $-1.616\pm0.252$ & $-1.597\pm0.252$ & $0.1143$ & \mbox{pop} \\
		GW191103\_012549 & $11.79$ & $7.88$ & $0.464$ & $0.495$ & $0.211$ & $0.803$ & $0.876$ & $-4.443\pm0.111$ & $-4.422\pm0.111$ & $-4.454\pm0.111$ & $0.0077$ & \mbox{pop} \\
		GW191105\_143521 & $10.67$ & $7.66$ & $0.231$ & $0.339$ & $-0.018$ & $0.743$ & $0.847$ & $+0.855\pm0.111$ & $+0.802\pm0.111$ & $+0.887\pm0.111$ & $0.5833$ & \mbox{acc weak} \\
		GW191109\_010717 & $65.11$ & $46.90$ & $0.826$ & $0.650$ & $-0.293$ & $0.729$ & $0.901$ & $+1.608\pm0.014$ & $+0.354\pm0.011$ & $+1.457\pm0.013$ & $0.7427$ & \mbox{acc} \\
		GW191127\_050227 & $53.16$ & $24.02$ & $0.659$ & $0.538$ & $0.179$ & $0.890$ & $0.899$ & $+0.443\pm0.059$ & $+0.186\pm0.058$ & $+0.406\pm0.059$ & $0.4854$ & \mbox{no pref} \\
		GW191129\_134029 & $10.68$ & $6.75$ & $0.248$ & $0.344$ & $0.064$ & $0.595$ & $0.827$ & $+0.011\pm0.160$ & $-0.039\pm0.160$ & $+0.038\pm0.160$ & $0.3839$ & \mbox{no pref} \\
		GW191204\_171526 & $11.90$ & $8.22$ & $0.396$ & $0.457$ & $0.159$ & $0.724$ & $0.809$ & $-4.467\pm0.057$ & $-4.468\pm0.057$ & $-4.457\pm0.057$ & $0.0075$ & \mbox{pop} \\
		GW191215\_223052 & $24.95$ & $18.12$ & $0.472$ & $0.444$ & $-0.041$ & $0.872$ & $0.887$ & $+0.152\pm0.072$ & $+0.145\pm0.072$ & $+0.159\pm0.071$ & $0.4165$ & \mbox{no pref} \\
		GW191216\_213338 & $12.07$ & $7.67$ & $0.241$ & $0.358$ & $0.111$ & $0.577$ & $0.823$ & $-2.991\pm0.081$ & $-3.043\pm0.081$ & $-2.951\pm0.081$ & $0.0318$ & \mbox{pop} \\
		GW191222\_033537 & $45.07$ & $34.72$ & $0.379$ & $0.413$ & $-0.041$ & $0.845$ & $0.876$ & $-0.190\pm0.024$ & $-0.081\pm0.023$ & $-0.163\pm0.023$ & $0.3392$ & \mbox{no pref} \\
		GW191230\_180458 & $49.40$ & $36.72$ & $0.510$ & $0.491$ & $-0.049$ & $0.900$ & $0.887$ & $+0.086\pm0.029$ & $+0.057\pm0.027$ & $+0.084\pm0.028$ & $0.4010$ & \mbox{no pref} \\
		GW200112\_155838 & $35.57$ & $28.30$ & $0.345$ & $0.386$ & $0.065$ & $0.771$ & $0.846$ & $+0.102\pm0.034$ & $+0.077\pm0.035$ & $+0.139\pm0.034$ & $0.4048$ & \mbox{no pref} \\
		GW200128\_022011 & $42.17$ & $32.63$ & $0.585$ & $0.503$ & $0.121$ & $0.889$ & $0.888$ & $+0.181\pm0.031$ & $+0.099\pm0.030$ & $+0.162\pm0.030$ & $0.4232$ & \mbox{no pref} \\
		GW200129\_065458 & $34.51$ & $28.94$ & $0.533$ & $0.489$ & $0.112$ & $0.892$ & $0.861$ & $+0.013\pm0.054$ & $-0.026\pm0.055$ & $+0.020\pm0.053$ & $0.3844$ & \mbox{no pref} \\
		GW200202\_154313 & $10.15$ & $7.34$ & $0.222$ & $0.331$ & $0.039$ & $0.647$ & $0.830$ & $+0.800\pm0.083$ & $+0.746\pm0.083$ & $+0.826\pm0.083$ & $0.5706$ & \mbox{acc weak} \\
		GW200208\_130117 & $37.82$ & $27.42$ & $0.356$ & $0.428$ & $-0.068$ & $0.836$ & $0.878$ & $-0.014\pm0.021$ & $-0.002\pm0.021$ & $+0.015\pm0.021$ & $0.3783$ & \mbox{no pref} \\
		GW200209\_085452 & $35.61$ & $27.14$ & $0.515$ & $0.490$ & $-0.117$ & $0.900$ & $0.890$ & $+0.001\pm0.035$ & $-0.003\pm0.035$ & $-0.002\pm0.035$ & $0.3817$ & \mbox{no pref} \\
		GW200216\_220804 & $51.29$ & $30.06$ & $0.484$ & $0.507$ & $0.096$ & $0.894$ & $0.893$ & $+0.080\pm0.048$ & $+0.049\pm0.047$ & $+0.074\pm0.048$ & $0.3998$ & \mbox{no pref} \\
		GW200219\_094415 & $37.53$ & $27.95$ & $0.474$ & $0.482$ & $-0.080$ & $0.886$ & $0.893$ & $-0.019\pm0.028$ & $-0.018\pm0.028$ & $-0.016\pm0.027$ & $0.3770$ & \mbox{no pref} \\
		GW200224\_222234 & $40.01$ & $32.54$ & $0.459$ & $0.435$ & $0.099$ & $0.855$ & $0.872$ & $+0.075\pm0.020$ & $+0.073\pm0.020$ & $+0.085\pm0.020$ & $0.3984$ & \mbox{no pref} \\
		GW200225\_060421 & $19.33$ & $13.99$ & $0.587$ & $0.425$ & $-0.116$ & $0.862$ & $0.880$ & $+0.176\pm0.134$ & $+0.190\pm0.134$ & $+0.158\pm0.134$ & $0.4219$ & \mbox{no pref} \\
		GW200302\_015811 & $37.77$ & $19.98$ & $0.372$ & $0.446$ & $0.014$ & $0.841$ & $0.887$ & $+0.021\pm0.027$ & $+0.013\pm0.027$ & $+0.044\pm0.027$ & $0.3860$ & \mbox{no pref} \\
		GW200311\_115853 & $34.23$ & $27.72$ & $0.393$ & $0.407$ & $-0.019$ & $0.839$ & $0.875$ & $+0.109\pm0.017$ & $+0.083\pm0.017$ & $+0.136\pm0.017$ & $0.4063$ & \mbox{no pref} \\
		GW200316\_215756 & $13.14$ & $7.81$ & $0.324$ & $0.438$ & $0.126$ & $0.650$ & $0.853$ & $-1.262\pm0.125$ & $-1.289\pm0.125$ & $-1.234\pm0.125$ & $0.1534$ & \mbox{pop} \\
		GW230601\_224134 & $63.56$ & $43.97$ & $0.409$ & $0.466$ & $-0.028$ & $0.883$ & $0.883$ & $-0.302\pm0.019$ & $-0.094\pm0.018$ & $-0.283\pm0.019$ & $0.3155$ & \mbox{no pref} \\
		GW230605\_065343 & $17.36$ & $11.00$ & $0.269$ & $0.415$ & $0.065$ & $0.741$ & $0.877$ & $+0.445\pm0.072$ & $+0.404\pm0.072$ & $+0.491\pm0.072$ & $0.4859$ & \mbox{no pref} \\
		GW230606\_004305 & $37.57$ & $25.94$ & $0.491$ & $0.497$ & $-0.099$ & $0.886$ & $0.890$ & $-0.028\pm0.025$ & $-0.026\pm0.025$ & $-0.032\pm0.025$ & $0.3750$ & \mbox{no pref} \\
		GW230608\_205047 & $48.22$ & $30.50$ & $0.388$ & $0.447$ & $0.037$ & $0.870$ & $0.886$ & $-0.179\pm0.020$ & $-0.058\pm0.019$ & $-0.157\pm0.020$ & $0.3417$ & \mbox{no pref} \\
		GW230609\_064958 & $35.36$ & $25.17$ & $0.449$ & $0.457$ & $-0.139$ & $0.884$ & $0.887$ & $+0.009\pm0.025$ & $+0.007\pm0.025$ & $+0.020\pm0.024$ & $0.3834$ & \mbox{no pref} \\
		GW230624\_113103 & $27.11$ & $16.11$ & $0.426$ & $0.487$ & $0.166$ & $0.849$ & $0.892$ & $+0.210\pm0.045$ & $+0.189\pm0.045$ & $+0.219\pm0.045$ & $0.4301$ & \mbox{no pref} \\
		GW230627\_015337 & $8.45$ & $5.74$ & $0.131$ & $0.230$ & $0.023$ & $0.388$ & $0.707$ & $+1.647\pm0.063$ & $+1.573\pm0.063$ & $+1.653\pm0.063$ & $0.7498$ & \mbox{acc} \\
		GW230628\_231200 & $32.57$ & $27.40$ & $0.343$ & $0.394$ & $-0.014$ & $0.853$ & $0.869$ & $+0.187\pm0.020$ & $+0.151\pm0.020$ & $+0.227\pm0.020$ & $0.4246$ & \mbox{no pref} \\
		GW230630\_125806 & $50.61$ & $32.75$ & $0.543$ & $0.555$ & $0.165$ & $0.879$ & $0.893$ & $+0.191\pm0.024$ & $+0.067\pm0.023$ & $+0.162\pm0.024$ & $0.4254$ & \mbox{no pref} \\
		GW230630\_234532 & $10.04$ & $6.64$ & $0.244$ & $0.384$ & $-0.039$ & $0.736$ & $0.864$ & $+0.608\pm0.113$ & $+0.562\pm0.113$ & $+0.633\pm0.113$ & $0.5250$ & \mbox{acc weak} \\
		GW230702\_185453 & $41.80$ & $17.14$ & $0.337$ & $0.489$ & $0.059$ & $0.828$ & $0.893$ & $-0.128\pm0.037$ & $-0.063\pm0.037$ & $-0.107\pm0.037$ & $0.3526$ & \mbox{no pref} \\
		GW230704\_021211 & $32.44$ & $20.12$ & $0.286$ & $0.416$ & $-0.009$ & $0.779$ & $0.880$ & $+0.131\pm0.031$ & $+0.113\pm0.031$ & $+0.172\pm0.031$ & $0.4115$ & \mbox{no pref} \\
		GW230704\_212616 & $88.89$ & $50.28$ & $0.644$ & $0.521$ & $0.273$ & $0.871$ & $0.891$ & $+0.601\pm0.015$ & $+0.291\pm0.013$ & $+0.564\pm0.014$ & $0.5232$ & \mbox{acc weak} \\
		GW230706\_104333 & $16.35$ & $11.51$ & $0.480$ & $0.528$ & $0.187$ & $0.862$ & $0.884$ & $-1.704\pm0.265$ & $-1.688\pm0.265$ & $-1.719\pm0.265$ & $0.1051$ & \mbox{pop} \\
		GW230707\_124047 & $45.75$ & $36.35$ & $0.443$ & $0.456$ & $-0.055$ & $0.881$ & $0.887$ & $-0.083\pm0.018$ & $-0.033\pm0.017$ & $-0.071\pm0.018$ & $0.3627$ & \mbox{no pref} \\
		GW230708\_053705 & $28.98$ & $22.61$ & $0.508$ & $0.483$ & $0.054$ & $0.880$ & $0.883$ & $+0.013\pm0.043$ & $+0.013\pm0.043$ & $+0.012\pm0.043$ & $0.3843$ & \mbox{no pref} \\
		GW230708\_230935 & $63.23$ & $39.42$ & $0.357$ & $0.460$ & $0.007$ & $0.852$ & $0.884$ & $-0.434\pm0.020$ & $-0.169\pm0.019$ & $-0.411\pm0.020$ & $0.2885$ & \mbox{no pref} \\
		GW230709\_122727 & $44.61$ & $29.61$ & $0.448$ & $0.482$ & $0.073$ & $0.873$ & $0.884$ & $+0.005\pm0.025$ & $+0.035\pm0.025$ & $+0.010\pm0.025$ & $0.3825$ & \mbox{no pref} \\
		GW230712\_090405 & $32.22$ & $12.51$ & $0.654$ & $0.520$ & $-0.025$ & $0.836$ & $0.891$ & $-0.061\pm0.041$ & $-0.090\pm0.041$ & $-0.100\pm0.041$ & $0.3677$ & \mbox{no pref} \\
		GW230723\_101834 & $16.63$ & $10.63$ & $0.395$ & $0.503$ & $-0.177$ & $0.847$ & $0.882$ & $-0.431\pm0.110$ & $-0.434\pm0.110$ & $-0.426\pm0.110$ & $0.2892$ & \mbox{no pref} \\
		GW230726\_002940 & $35.43$ & $27.97$ & $0.382$ & $0.429$ & $-0.017$ & $0.855$ & $0.872$ & $+0.035\pm0.023$ & $+0.022\pm0.023$ & $+0.063\pm0.023$ & $0.3894$ & \mbox{no pref} \\
		GW230729\_082317 & $12.29$ & $7.61$ & $0.356$ & $0.461$ & $0.116$ & $0.772$ & $0.880$ & $-0.177\pm0.204$ & $-0.199\pm0.204$ & $-0.156\pm0.204$ & $0.3421$ & \mbox{no pref} \\
		GW230731\_215307 & $10.32$ & $7.86$ & $0.281$ & $0.361$ & $-0.050$ & $0.795$ & $0.854$ & $+0.490\pm0.121$ & $+0.445\pm0.121$ & $+0.517\pm0.121$ & $0.4968$ & \mbox{no pref} \\
		GW230803\_033412 & $44.49$ & $28.52$ & $0.528$ & $0.472$ & $0.066$ & $0.883$ & $0.886$ & $+0.148\pm0.017$ & $+0.074\pm0.017$ & $+0.146\pm0.017$ & $0.4155$ & \mbox{no pref} \\
		GW230805\_034249 & $31.47$ & $22.91$ & $0.442$ & $0.461$ & $0.047$ & $0.884$ & $0.886$ & $+0.150\pm0.036$ & $+0.134\pm0.036$ & $+0.163\pm0.036$ & $0.4159$ & \mbox{no pref} \\
		GW230806\_204041 & $50.71$ & $34.94$ & $0.416$ & $0.462$ & $0.069$ & $0.875$ & $0.887$ & $-0.160\pm0.020$ & $-0.046\pm0.019$ & $-0.145\pm0.020$ & $0.3458$ & \mbox{no pref} \\
		GW230811\_032116 & $35.06$ & $22.70$ & $0.380$ & $0.437$ & $0.023$ & $0.830$ & $0.882$ & $+0.107\pm0.022$ & $+0.087\pm0.023$ & $+0.131\pm0.022$ & $0.4058$ & \mbox{no pref} \\
		GW230814\_061920 & $69.85$ & $42.54$ & $0.402$ & $0.471$ & $0.068$ & $0.858$ & $0.888$ & $-0.352\pm0.019$ & $-0.151\pm0.018$ & $-0.336\pm0.019$ & $0.3052$ & \mbox{no pref} \\
		GW230814\_230901 & $33.84$ & $27.98$ & $0.160$ & $0.174$ & $-0.006$ & $0.447$ & $0.608$ & $+0.527\pm0.014$ & $+0.449\pm0.014$ & $+0.616\pm0.014$ & $0.5056$ & \mbox{acc weak} \\
		GW230819\_171910 & $69.98$ & $35.14$ & $0.565$ & $0.504$ & $-0.037$ & $0.895$ & $0.893$ & $+0.233\pm0.018$ & $+0.132\pm0.017$ & $+0.225\pm0.018$ & $0.4354$ & \mbox{no pref} \\
		GW230820\_212515 & $61.75$ & $34.38$ & $0.508$ & $0.535$ & $0.212$ & $0.880$ & $0.893$ & $+0.093\pm0.017$ & $+0.024\pm0.015$ & $+0.081\pm0.016$ & $0.4028$ & \mbox{no pref} \\
		GW230824\_033047 & $52.56$ & $35.93$ & $0.387$ & $0.427$ & $-0.008$ & $0.868$ & $0.881$ & $-0.238\pm0.021$ & $-0.055\pm0.020$ & $-0.212\pm0.021$ & $0.3290$ & \mbox{no pref} \\
		GW230825\_041334 & $43.32$ & $26.82$ & $0.697$ & $0.509$ & $0.278$ & $0.828$ & $0.890$ & $+0.394\pm0.026$ & $+0.204\pm0.025$ & $+0.331\pm0.026$ & $0.4737$ & \mbox{no pref} \\
		GW230831\_015414 & $42.27$ & $30.38$ & $0.519$ & $0.491$ & $0.035$ & $0.899$ & $0.893$ & $+0.044\pm0.016$ & $+0.013\pm0.015$ & $+0.045\pm0.016$ & $0.3913$ & \mbox{no pref} \\
		GW230904\_051013 & $10.57$ & $7.11$ & $0.278$ & $0.375$ & $0.051$ & $0.733$ & $0.869$ & $+0.335\pm0.131$ & $+0.290\pm0.131$ & $+0.359\pm0.131$ & $0.4597$ & \mbox{no pref} \\
		GW230911\_195324 & $33.89$ & $21.59$ & $0.438$ & $0.452$ & $-0.021$ & $0.876$ & $0.883$ & $+0.091\pm0.020$ & $+0.072\pm0.020$ & $+0.108\pm0.020$ & $0.4023$ & \mbox{no pref} \\
		GW230914\_111401 & $58.22$ & $38.08$ & $0.318$ & $0.540$ & $0.129$ & $0.813$ & $0.871$ & $-0.430\pm0.018$ & $-0.237\pm0.017$ & $-0.423\pm0.018$ & $0.2894$ & \mbox{no pref} \\
		GW230919\_215712 & $27.35$ & $21.43$ & $0.447$ & $0.475$ & $0.184$ & $0.838$ & $0.873$ & $-0.430\pm0.029$ & $-0.426\pm0.029$ & $-0.435\pm0.029$ & $0.2893$ & \mbox{no pref} \\
		GW230920\_071124 & $32.51$ & $24.09$ & $0.410$ & $0.464$ & $0.006$ & $0.874$ & $0.889$ & $+0.089\pm0.028$ & $+0.069\pm0.028$ & $+0.107\pm0.027$ & $0.4017$ & \mbox{no pref} \\
		GW230922\_020344 & $38.65$ & $29.42$ & $0.342$ & $0.427$ & $0.021$ & $0.832$ & $0.881$ & $-0.036\pm0.019$ & $-0.030\pm0.019$ & $-0.004\pm0.018$ & $0.3732$ & \mbox{no pref} \\
		GW230922\_040658 & $75.17$ & $51.01$ & $0.628$ & $0.532$ & $0.299$ & $0.841$ & $0.884$ & $+0.635\pm0.019$ & $+0.240\pm0.017$ & $+0.580\pm0.019$ & $0.5313$ & \mbox{acc weak} \\
		GW230924\_124453 & $28.71$ & $23.26$ & $0.310$ & $0.376$ & $0.026$ & $0.819$ & $0.859$ & $+0.345\pm0.026$ & $+0.305\pm0.026$ & $+0.390\pm0.026$ & $0.4620$ & \mbox{no pref} \\
		GW230927\_043729 & $34.70$ & $27.07$ & $0.341$ & $0.404$ & $0.001$ & $0.849$ & $0.876$ & $+0.099\pm0.020$ & $+0.079\pm0.020$ & $+0.134\pm0.020$ & $0.4042$ & \mbox{no pref} \\
		GW230927\_153832 & $21.71$ & $16.64$ & $0.278$ & $0.341$ & $0.032$ & $0.742$ & $0.841$ & $+0.586\pm0.033$ & $+0.535\pm0.033$ & $+0.642\pm0.033$ & $0.5196$ & \mbox{acc weak} \\
		GW230928\_215827 & $53.52$ & $29.68$ & $0.736$ & $0.491$ & $0.381$ & $0.723$ & $0.891$ & $+0.820\pm0.023$ & $+0.317\pm0.021$ & $+0.728\pm0.022$ & $0.5752$ & \mbox{acc weak} \\
		GW230930\_110730 & $34.25$ & $24.54$ & $0.375$ & $0.437$ & $0.018$ & $0.857$ & $0.878$ & $+0.076\pm0.030$ & $+0.072\pm0.030$ & $+0.100\pm0.030$ & $0.3986$ & \mbox{no pref} \\
		GW231001\_140220 & $75.63$ & $40.51$ & $0.407$ & $0.484$ & $-0.037$ & $0.873$ & $0.889$ & $-0.328\pm0.017$ & $-0.158\pm0.016$ & $-0.314\pm0.017$ & $0.3101$ & \mbox{no pref} \\
		GW231004\_232346 & $64.45$ & $34.74$ & $0.445$ & $0.489$ & $-0.044$ & $0.882$ & $0.891$ & $-0.133\pm0.016$ & $-0.039\pm0.015$ & $-0.126\pm0.016$ & $0.3517$ & \mbox{no pref} \\
		GW231005\_021030 & $83.69$ & $49.78$ & $0.556$ & $0.491$ & $0.101$ & $0.887$ & $0.888$ & $+0.198\pm0.015$ & $+0.127\pm0.013$ & $+0.189\pm0.015$ & $0.4272$ & \mbox{no pref} \\
		GW231005\_091549 & $28.64$ & $21.20$ & $0.407$ & $0.454$ & $-0.028$ & $0.881$ & $0.890$ & $+0.143\pm0.036$ & $+0.125\pm0.036$ & $+0.165\pm0.036$ & $0.4144$ & \mbox{no pref} \\
		GW231008\_142521 & $45.19$ & $25.76$ & $0.446$ & $0.467$ & $-0.004$ & $0.872$ & $0.888$ & $-0.018\pm0.024$ & $+0.019\pm0.023$ & $-0.007\pm0.023$ & $0.3773$ & \mbox{no pref} \\
		GW231014\_040532 & $20.65$ & $14.73$ & $0.534$ & $0.502$ & $0.175$ & $0.878$ & $0.883$ & $-0.466\pm0.090$ & $-0.454\pm0.090$ & $-0.482\pm0.090$ & $0.2822$ & \mbox{no pref} \\
		GW231018\_233037 & $11.55$ & $7.27$ & $0.255$ & $0.411$ & $0.009$ & $0.757$ & $0.880$ & $+0.853\pm0.098$ & $+0.811\pm0.098$ & $+0.885\pm0.098$ & $0.5829$ & \mbox{acc weak} \\
		GW231020\_142947 & $12.03$ & $7.37$ & $0.400$ & $0.442$ & $0.144$ & $0.746$ & $0.876$ & $-1.872\pm0.287$ & $-1.890\pm0.287$ & $-1.858\pm0.286$ & $0.0905$ & \mbox{pop} \\
		GW231026\_130704 & $32.44$ & $21.41$ & $0.472$ & $0.485$ & $0.012$ & $0.886$ & $0.887$ & $+0.050\pm0.039$ & $+0.025\pm0.039$ & $+0.054\pm0.039$ & $0.3927$ & \mbox{no pref} \\
		GW231028\_153006 & $94.85$ & $54.11$ & $0.744$ & $0.540$ & $0.388$ & $0.511$ & $0.883$ & $+1.625\pm0.016$ & $+0.564\pm0.014$ & $+1.449\pm0.015$ & $0.7459$ & \mbox{acc} \\
		GW231029\_111508 & $63.78$ & $42.63$ & $0.414$ & $0.523$ & $0.145$ & $0.871$ & $0.885$ & $-0.173\pm0.019$ & $-0.115\pm0.018$ & $-0.177\pm0.019$ & $0.3429$ & \mbox{no pref} \\
		GW231102\_071736 & $60.93$ & $42.10$ & $0.348$ & $0.439$ & $0.059$ & $0.845$ & $0.887$ & $-0.473\pm0.018$ & $-0.160\pm0.017$ & $-0.444\pm0.018$ & $0.2808$ & \mbox{no pref} \\
		GW231104\_133418 & $12.20$ & $8.62$ & $0.357$ & $0.468$ & $0.140$ & $0.777$ & $0.865$ & $-3.542\pm0.155$ & $-3.557\pm0.155$ & $-3.520\pm0.155$ & $0.0186$ & \mbox{pop} \\
		GW231108\_125142 & $23.23$ & $17.41$ & $0.313$ & $0.423$ & $-0.081$ & $0.833$ & $0.879$ & $+0.306\pm0.035$ & $+0.273\pm0.035$ & $+0.342\pm0.035$ & $0.4526$ & \mbox{no pref} \\
		GW231110\_040320 & $19.39$ & $12.61$ & $0.452$ & $0.521$ & $0.166$ & $0.820$ & $0.882$ & $-1.715\pm0.099$ & $-1.704\pm0.099$ & $-1.726\pm0.099$ & $0.1041$ & \mbox{pop} \\
		GW231113\_122623 & $39.81$ & $26.42$ & $0.701$ & $0.543$ & $0.331$ & $0.841$ & $0.890$ & $+0.200\pm0.023$ & $+0.069\pm0.023$ & $+0.131\pm0.023$ & $0.4276$ & \mbox{no pref} \\
		GW231113\_150041 & $56.30$ & $29.33$ & $0.681$ & $0.499$ & $0.035$ & $0.880$ & $0.892$ & $+0.536\pm0.028$ & $+0.248\pm0.026$ & $+0.509\pm0.027$ & $0.5077$ & \mbox{acc weak} \\
		GW231113\_200417 & $11.58$ & $7.38$ & $0.413$ & $0.468$ & $0.126$ & $0.813$ & $0.882$ & $-3.245\pm0.305$ & $-3.254\pm0.305$ & $-3.239\pm0.305$ & $0.0249$ & \mbox{pop} \\
		GW231114\_043211 & $22.76$ & $8.19$ & $0.210$ & $0.450$ & $0.081$ & $0.485$ & $0.883$ & $+0.584\pm0.051$ & $+0.541\pm0.051$ & $+0.631\pm0.051$ & $0.5193$ & \mbox{acc weak} \\
		GW231118\_005626 & $20.01$ & $10.84$ & $0.660$ & $0.565$ & $0.379$ & $0.685$ & $0.893$ & $-3.905\pm0.124$ & $-3.778\pm0.124$ & $-4.036\pm0.124$ & $0.0130$ & \mbox{pop} \\
		GW231118\_071402 & $42.61$ & $29.01$ & $0.461$ & $0.486$ & $0.110$ & $0.875$ & $0.889$ & $-0.008\pm0.022$ & $+0.014\pm0.022$ & $-0.006\pm0.022$ & $0.3796$ & \mbox{no pref} \\
		GW231118\_090602 & $12.92$ & $7.36$ & $0.391$ & $0.420$ & $0.080$ & $0.786$ & $0.875$ & $+0.024\pm0.161$ & $-0.003\pm0.161$ & $+0.041\pm0.161$ & $0.3868$ & \mbox{no pref} \\
		GW231119\_075248 & $48.64$ & $33.34$ & $0.395$ & $0.456$ & $-0.007$ & $0.872$ & $0.882$ & $-0.197\pm0.022$ & $-0.090\pm0.021$ & $-0.179\pm0.022$ & $0.3376$ & \mbox{no pref} \\
		GW231123\_135430 & $127.84$ & $107.70$ & $0.903$ & $0.905$ & $0.270$ & $0.355$ & $0.394$ & $+4.782\pm0.018$ & $+1.811\pm0.013$ & $+4.473\pm0.017$ & $0.9850$ & \mbox{acc} \\
		GW231127\_165300 & $45.16$ & $28.71$ & $0.432$ & $0.485$ & $0.046$ & $0.876$ & $0.889$ & $-0.049\pm0.025$ & $-0.003\pm0.025$ & $-0.040\pm0.025$ & $0.3702$ & \mbox{no pref} \\
		GW231129\_081745 & $44.91$ & $23.66$ & $0.376$ & $0.468$ & $0.028$ & $0.867$ & $0.889$ & $-0.144\pm0.028$ & $-0.068\pm0.028$ & $-0.124\pm0.028$ & $0.3491$ & \mbox{no pref} \\
		GW231206\_233134 & $35.73$ & $28.10$ & $0.525$ & $0.470$ & $-0.087$ & $0.895$ & $0.898$ & $+0.074\pm0.023$ & $+0.055\pm0.023$ & $+0.074\pm0.022$ & $0.3983$ & \mbox{no pref} \\
		GW231206\_233901 & $37.19$ & $28.80$ & $0.440$ & $0.408$ & $-0.053$ & $0.858$ & $0.858$ & $+0.076\pm0.019$ & $+0.071\pm0.020$ & $+0.094\pm0.019$ & $0.3988$ & \mbox{no pref} \\
		GW231213\_111417 & $35.60$ & $27.18$ & $0.385$ & $0.425$ & $0.051$ & $0.851$ & $0.879$ & $+0.057\pm0.022$ & $+0.047\pm0.022$ & $+0.081\pm0.022$ & $0.3943$ & \mbox{no pref} \\
		GW231221\_135041 & $46.78$ & $28.98$ & $0.605$ & $0.498$ & $0.013$ & $0.884$ & $0.891$ & $+0.267\pm0.019$ & $+0.118\pm0.018$ & $+0.248\pm0.019$ & $0.4435$ & \mbox{no pref} \\
		GW231223\_032836 & $45.40$ & $31.38$ & $0.524$ & $0.528$ & $-0.164$ & $0.890$ & $0.886$ & $-0.035\pm0.024$ & $-0.040\pm0.023$ & $-0.047\pm0.024$ & $0.3735$ & \mbox{no pref} \\
		GW231223\_075055 & $11.94$ & $6.79$ & $0.416$ & $0.432$ & $0.065$ & $0.815$ & $0.882$ & $+0.255\pm0.151$ & $+0.235\pm0.151$ & $+0.265\pm0.151$ & $0.4407$ & \mbox{no pref} \\
		GW231223\_202619 & $11.14$ & $8.32$ & $0.334$ & $0.458$ & $0.097$ & $0.798$ & $0.875$ & $-0.307\pm0.207$ & $-0.332\pm0.207$ & $-0.284\pm0.207$ & $0.3144$ & \mbox{no pref} \\
		GW231224\_024321 & $9.31$ & $7.31$ & $0.292$ & $0.318$ & $-0.008$ & $0.807$ & $0.829$ & $+1.131\pm0.096$ & $+1.082\pm0.096$ & $+1.145\pm0.096$ & $0.6460$ & \mbox{acc} \\
		GW231226\_101520 & $40.16$ & $35.15$ & $0.211$ & $0.422$ & $-0.065$ & $0.593$ & $0.845$ & $-0.253\pm0.016$ & $-0.263\pm0.016$ & $-0.208\pm0.016$ & $0.3257$ & \mbox{no pref} \\
		GW231230\_170116 & $53.37$ & $34.74$ & $0.566$ & $0.517$ & $-0.185$ & $0.885$ & $0.894$ & $+0.196\pm0.018$ & $+0.100\pm0.017$ & $+0.178\pm0.017$ & $0.4268$ & \mbox{no pref} \\
		GW231231\_154016 & $22.49$ & $17.21$ & $0.295$ & $0.372$ & $-0.032$ & $0.816$ & $0.865$ & $+0.520\pm0.033$ & $+0.475\pm0.033$ & $+0.568\pm0.033$ & $0.5038$ & \mbox{acc weak} \\
		GW240104\_164932 & $42.16$ & $31.70$ & $0.499$ & $0.448$ & $0.076$ & $0.852$ & $0.872$ & $+0.167\pm0.020$ & $+0.166\pm0.019$ & $+0.161\pm0.019$ & $0.4198$ & \mbox{no pref} \\
		GW240107\_013215 & $58.51$ & $33.17$ & $0.708$ & $0.564$ & $0.323$ & $0.821$ & $0.895$ & $+0.838\pm0.019$ & $+0.298\pm0.018$ & $+0.760\pm0.019$ & $0.5794$ & \mbox{acc weak} \\
		GW240109\_050431 & $28.74$ & $18.14$ & $0.313$ & $0.438$ & $-0.074$ & $0.815$ & $0.879$ & $+0.157\pm0.026$ & $+0.129\pm0.026$ & $+0.190\pm0.026$ & $0.4177$ & \mbox{no pref} \\
		GW240413\_022019 & $7.92$ & $5.57$ & $0.256$ & $0.363$ & $0.066$ & $0.687$ & $0.862$ & $-0.942\pm0.391$ & $-0.991\pm0.391$ & $-0.940\pm0.391$ & $0.1985$ & \mbox{pop} \\
		GW240414\_054515 & $38.42$ & $25.83$ & $0.314$ & $0.424$ & $0.009$ & $0.823$ & $0.884$ & $-0.040\pm0.021$ & $-0.022\pm0.021$ & $-0.003\pm0.021$ & $0.3724$ & \mbox{no pref} \\
		GW240420\_175625 & $34.64$ & $23.87$ & $0.301$ & $0.416$ & $-0.007$ & $0.820$ & $0.875$ & $+0.055\pm0.026$ & $+0.040\pm0.026$ & $+0.093\pm0.026$ & $0.3939$ & \mbox{no pref} \\
		GW240426\_031451 & $51.08$ & $36.53$ & $0.407$ & $0.463$ & $-0.048$ & $0.890$ & $0.886$ & $-0.186\pm0.020$ & $-0.059\pm0.019$ & $-0.168\pm0.020$ & $0.3401$ & \mbox{no pref} \\
		GW240428\_225440 & $19.77$ & $14.66$ & $0.240$ & $0.331$ & $0.032$ & $0.738$ & $0.850$ & $+1.003\pm0.049$ & $+0.949\pm0.049$ & $+1.063\pm0.049$ & $0.6174$ & \mbox{acc} \\
		GW240501\_033534 & $38.51$ & $28.10$ & $0.434$ & $0.481$ & $0.028$ & $0.872$ & $0.889$ & $+0.021\pm0.021$ & $+0.023\pm0.021$ & $+0.031\pm0.021$ & $0.3861$ & \mbox{no pref} \\
		GW240505\_133552 & $31.01$ & $23.38$ & $0.471$ & $0.482$ & $-0.137$ & $0.884$ & $0.883$ & $-0.055\pm0.033$ & $-0.057\pm0.033$ & $-0.053\pm0.033$ & $0.3689$ & \mbox{no pref} \\
		GW240507\_041632 & $31.27$ & $8.53$ & $0.358$ & $0.465$ & $0.013$ & $0.813$ & $0.888$ & $+0.061\pm0.060$ & $+0.042\pm0.060$ & $+0.083\pm0.060$ & $0.3954$ & \mbox{no pref} \\
		GW240511\_031507 & $40.05$ & $32.08$ & $0.296$ & $0.373$ & $0.029$ & $0.803$ & $0.851$ & $-0.112\pm0.017$ & $-0.095\pm0.017$ & $-0.066\pm0.017$ & $0.3564$ & \mbox{no pref} \\
		GW240512\_024139 & $12.62$ & $7.98$ & $0.390$ & $0.496$ & $0.141$ & $0.798$ & $0.878$ & $-3.080\pm0.393$ & $-3.089\pm0.393$ & $-3.064\pm0.393$ & $0.0292$ & \mbox{pop} \\
		GW240513\_183302 & $24.35$ & $16.90$ & $0.577$ & $0.498$ & $0.138$ & $0.830$ & $0.884$ & $-1.249\pm0.068$ & $-1.212\pm0.068$ & $-1.288\pm0.068$ & $0.1551$ & \mbox{pop} \\
		GW240514\_121713 & $46.12$ & $36.63$ & $0.296$ & $0.353$ & $-0.005$ & $0.823$ & $0.849$ & $-0.386\pm0.015$ & $-0.199\pm0.015$ & $-0.338\pm0.015$ & $0.2982$ & \mbox{no pref} \\
		GW240515\_005301 & $36.85$ & $18.51$ & $0.709$ & $0.558$ & $0.449$ & $0.623$ & $0.890$ & $+0.174\pm0.056$ & $+0.141\pm0.056$ & $+0.027\pm0.056$ & $0.4215$ & \mbox{no pref} \\
		GW240519\_012815 & $63.72$ & $40.13$ & $0.394$ & $0.476$ & $-0.074$ & $0.869$ & $0.888$ & $-0.303\pm0.020$ & $-0.109\pm0.019$ & $-0.288\pm0.020$ & $0.3151$ & \mbox{no pref} \\
		GW240520\_213616 & $11.43$ & $7.98$ & $0.228$ & $0.372$ & $-0.002$ & $0.734$ & $0.866$ & $+1.000\pm0.110$ & $+0.950\pm0.110$ & $+1.037\pm0.110$ & $0.6166$ & \mbox{acc weak} \\
		GW240525\_031210 & $31.07$ & $21.36$ & $0.525$ & $0.481$ & $0.166$ & $0.880$ & $0.885$ & $-0.050\pm0.035$ & $-0.049\pm0.035$ & $-0.055\pm0.035$ & $0.3701$ & \mbox{no pref} \\
		GW240526\_093944 & $14.50$ & $8.36$ & $0.580$ & $0.466$ & $0.217$ & $0.815$ & $0.882$ & $-2.623\pm0.406$ & $-2.588\pm0.406$ & $-2.659\pm0.406$ & $0.0452$ & \mbox{pop} \\
		GW240527\_183429 & $50.41$ & $33.78$ & $0.597$ & $0.529$ & $0.286$ & $0.866$ & $0.894$ & $+0.315\pm0.022$ & $+0.132\pm0.020$ & $+0.274\pm0.021$ & $0.4548$ & \mbox{no pref} \\
		GW240527\_230910 & $27.58$ & $9.97$ & $0.275$ & $0.483$ & $-0.031$ & $0.737$ & $0.890$ & $+0.168\pm0.061$ & $+0.142\pm0.061$ & $+0.197\pm0.061$ & $0.4202$ & \mbox{no pref} \\
		GW240530\_012417 & $14.10$ & $8.31$ & $0.308$ & $0.588$ & $0.043$ & $0.778$ & $0.893$ & $+0.319\pm0.184$ & $+0.314\pm0.184$ & $+0.340\pm0.184$ & $0.4558$ & \mbox{no pref} \\
		GW240531\_040326 & $19.52$ & $14.33$ & $0.566$ & $0.467$ & $0.144$ & $0.882$ & $0.886$ & $-0.077\pm0.129$ & $-0.067\pm0.129$ & $-0.088\pm0.129$ & $0.3640$ & \mbox{no pref} \\
		GW240531\_075248 & $32.95$ & $22.46$ & $0.463$ & $0.488$ & $0.024$ & $0.881$ & $0.890$ & $+0.052\pm0.030$ & $+0.042\pm0.030$ & $+0.059\pm0.030$ & $0.3933$ & \mbox{no pref} \\
		GW240601\_061200 & $54.60$ & $32.37$ & $0.454$ & $0.472$ & $-0.099$ & $0.875$ & $0.884$ & $-0.059\pm0.019$ & $+0.010\pm0.018$ & $-0.052\pm0.019$ & $0.3680$ & \mbox{no pref} \\
		GW240601\_231004 & $10.65$ & $7.69$ & $0.404$ & $0.461$ & $0.091$ & $0.851$ & $0.874$ & $-1.075\pm0.452$ & $-1.089\pm0.452$ & $-1.066\pm0.452$ & $0.1787$ & \mbox{pop} \\
		GW240612\_081540 & $54.68$ & $37.71$ & $0.503$ & $0.490$ & $0.071$ & $0.887$ & $0.887$ & $+0.044\pm0.024$ & $+0.045\pm0.023$ & $+0.042\pm0.024$ & $0.3914$ & \mbox{no pref} \\
		GW240615\_113620 & $33.89$ & $26.23$ & $0.337$ & $0.503$ & $-0.096$ & $0.796$ & $0.874$ & $-0.005\pm0.021$ & $-0.022\pm0.021$ & $+0.014\pm0.021$ & $0.3802$ & \mbox{no pref} \\
		GW240615\_160735 & $28.27$ & $20.30$ & $0.562$ & $0.571$ & $0.260$ & $0.864$ & $0.880$ & $-0.738\pm0.065$ & $-0.681\pm0.065$ & $-0.798\pm0.065$ & $0.2320$ & \mbox{pop} \\
		GW240618\_071627 & $63.33$ & $40.00$ & $0.396$ & $0.473$ & $-0.010$ & $0.875$ & $0.886$ & $-0.303\pm0.019$ & $-0.120\pm0.018$ & $-0.286\pm0.019$ & $0.3152$ & \mbox{no pref} \\
		GW240621\_195059 & $36.54$ & $29.48$ & $0.207$ & $0.266$ & $-0.012$ & $0.644$ & $0.778$ & $+0.085\pm0.015$ & $+0.027\pm0.015$ & $+0.157\pm0.015$ & $0.4008$ & \mbox{no pref} \\
		GW240621\_200935 & $41.32$ & $31.54$ & $0.428$ & $0.465$ & $-0.102$ & $0.875$ & $0.886$ & $-0.034\pm0.021$ & $-0.016\pm0.021$ & $-0.022\pm0.021$ & $0.3737$ & \mbox{no pref} \\
		GW240621\_214041 & $44.81$ & $31.71$ & $0.469$ & $0.462$ & $-0.007$ & $0.885$ & $0.885$ & $+0.010\pm0.020$ & $+0.019\pm0.020$ & $+0.018\pm0.020$ & $0.3836$ & \mbox{no pref} \\
		GW240622\_004008 & $18.56$ & $11.34$ & $0.615$ & $0.558$ & $0.246$ & $0.796$ & $0.891$ & $-3.482\pm0.131$ & $-3.401\pm0.131$ & $-3.563\pm0.131$ & $0.0197$ & \mbox{pop} \\
		GW240627\_131622 & $13.18$ & $6.72$ & $0.314$ & $0.430$ & $0.004$ & $0.779$ & $0.885$ & $+0.511\pm0.152$ & $+0.479\pm0.152$ & $+0.536\pm0.152$ & $0.5016$ & \mbox{acc weak} \\
		GW240629\_145256 & $11.05$ & $7.80$ & $0.231$ & $0.365$ & $0.044$ & $0.705$ & $0.868$ & $+1.070\pm0.132$ & $+1.019\pm0.132$ & $+1.106\pm0.132$ & $0.6325$ & \mbox{acc} \\
		GW240630\_101703 & $27.91$ & $20.89$ & $0.604$ & $0.514$ & $-0.109$ & $0.885$ & $0.896$ & $-0.086\pm0.047$ & $-0.069\pm0.048$ & $-0.105\pm0.047$ & $0.3621$ & \mbox{no pref} \\
		GW240703\_191355 & $34.17$ & $26.64$ & $0.360$ & $0.417$ & $0.051$ & $0.842$ & $0.876$ & $+0.123\pm0.021$ & $+0.098\pm0.022$ & $+0.154\pm0.021$ & $0.4096$ & \mbox{no pref} \\
		GW240705\_053215 & $47.11$ & $35.50$ & $0.317$ & $0.403$ & $-0.049$ & $0.829$ & $0.869$ & $-0.318\pm0.016$ & $-0.155\pm0.016$ & $-0.281\pm0.016$ & $0.3122$ & \mbox{no pref} \\
		GW240716\_034900 & $41.00$ & $26.32$ & $0.346$ & $0.470$ & $-0.033$ & $0.817$ & $0.887$ & $-0.101\pm0.019$ & $-0.054\pm0.019$ & $-0.077\pm0.019$ & $0.3588$ & \mbox{no pref} \\
		GW240824\_205609 & $68.02$ & $36.92$ & $0.442$ & $0.505$ & $0.068$ & $0.875$ & $0.894$ & $-0.127\pm0.024$ & $-0.051\pm0.022$ & $-0.124\pm0.023$ & $0.3530$ & \mbox{no pref} \\
		GW240825\_055146 & $13.20$ & $7.87$ & $0.391$ & $0.443$ & $0.073$ & $0.813$ & $0.888$ & $+0.458\pm0.158$ & $+0.431\pm0.158$ & $+0.480\pm0.158$ & $0.4890$ & \mbox{no pref} \\
		GW240830\_211120 & $11.87$ & $7.84$ & $0.248$ & $0.416$ & $0.067$ & $0.701$ & $0.873$ & $+0.414\pm0.160$ & $+0.371\pm0.160$ & $+0.450\pm0.160$ & $0.4785$ & \mbox{no pref} \\
		GW240902\_143306 & $24.44$ & $18.35$ & $0.519$ & $0.489$ & $-0.027$ & $0.887$ & $0.887$ & $-0.136\pm0.062$ & $-0.128\pm0.062$ & $-0.146\pm0.062$ & $0.3510$ & \mbox{no pref} \\
		GW240907\_153833 & $37.58$ & $28.82$ & $0.389$ & $0.434$ & $0.027$ & $0.854$ & $0.878$ & $-0.005\pm0.020$ & $+0.001\pm0.020$ & $+0.018\pm0.020$ & $0.3803$ & \mbox{no pref} \\
		GW240908\_082628 & $39.27$ & $27.68$ & $0.411$ & $0.455$ & $0.106$ & $0.864$ & $0.883$ & $+0.009\pm0.022$ & $+0.023\pm0.022$ & $+0.023\pm0.022$ & $0.3834$ & \mbox{no pref} \\
		GW240908\_125134 & $44.36$ & $32.48$ & $0.409$ & $0.444$ & $-0.094$ & $0.874$ & $0.886$ & $-0.127\pm0.019$ & $-0.045\pm0.018$ & $-0.109\pm0.018$ & $0.3530$ & \mbox{no pref} \\
		GW240910\_103535 & $10.65$ & $6.47$ & $0.255$ & $0.382$ & $0.108$ & $0.572$ & $0.857$ & $-3.162\pm0.084$ & $-3.209\pm0.084$ & $-3.137\pm0.084$ & $0.0269$ & \mbox{pop} \\
		GW240915\_001357 & $10.91$ & $7.74$ & $0.392$ & $0.446$ & $0.129$ & $0.742$ & $0.869$ & $-4.010\pm0.058$ & $-4.027\pm0.058$ & $-3.997\pm0.058$ & $0.0117$ & \mbox{pop} \\
		GW240915\_105151 & $11.64$ & $7.50$ & $0.402$ & $0.437$ & $0.072$ & $0.785$ & $0.883$ & $-0.284\pm0.196$ & $-0.303\pm0.196$ & $-0.271\pm0.196$ & $0.3193$ & \mbox{no pref} \\
		GW240916\_184352 & $10.68$ & $7.82$ & $0.348$ & $0.402$ & $0.054$ & $0.840$ & $0.873$ & $+0.089\pm0.227$ & $+0.055\pm0.227$ & $+0.108\pm0.227$ & $0.4017$ & \mbox{no pref} \\
		GW240919\_061559 & $38.04$ & $29.49$ & $0.299$ & $0.388$ & $-0.026$ & $0.788$ & $0.864$ & $-0.012\pm0.017$ & $-0.028\pm0.017$ & $+0.031\pm0.017$ & $0.3787$ & \mbox{no pref} \\
		GW240920\_073424 & $24.73$ & $13.37$ & $0.451$ & $0.609$ & $-0.091$ & $0.889$ & $0.888$ & $-0.850\pm0.103$ & $-0.812\pm0.103$ & $-0.892\pm0.103$ & $0.2131$ & \mbox{pop} \\
		GW240920\_124024 & $36.92$ & $32.46$ & $0.223$ & $0.639$ & $-0.036$ & $0.717$ & $0.846$ & $-0.363\pm0.018$ & $-0.356\pm0.018$ & $-0.370\pm0.017$ & $0.3028$ & \mbox{no pref} \\
		GW240921\_201835 & $34.86$ & $9.83$ & $0.199$ & $0.474$ & $0.005$ & $0.541$ & $0.889$ & $-0.098\pm0.048$ & $-0.126\pm0.048$ & $-0.053\pm0.048$ & $0.3593$ & \mbox{no pref} \\
		GW240922\_142106 & $11.45$ & $7.76$ & $0.297$ & $0.414$ & $0.073$ & $0.732$ & $0.877$ & $-1.109\pm0.313$ & $-1.147\pm0.313$ & $-1.080\pm0.313$ & $0.1739$ & \mbox{pop} \\
		GW240923\_204006 & $46.40$ & $35.62$ & $0.355$ & $0.432$ & $0.016$ & $0.844$ & $0.883$ & $-0.193\pm0.017$ & $-0.094\pm0.016$ & $-0.164\pm0.016$ & $0.3385$ & \mbox{no pref} \\
		GW240924\_000316 & $44.53$ & $33.65$ & $0.352$ & $0.408$ & $0.062$ & $0.844$ & $0.876$ & $-0.139\pm0.019$ & $-0.056\pm0.019$ & $-0.110\pm0.019$ & $0.3502$ & \mbox{no pref} \\
		GW240925\_005809 & $9.02$ & $6.99$ & $0.199$ & $0.266$ & $0.025$ & $0.615$ & $0.783$ & $+0.978\pm0.073$ & $+0.913\pm0.073$ & $+0.988\pm0.073$ & $0.6116$ & \mbox{acc weak} \\
		GW240930\_035959 & $25.53$ & $11.21$ & $0.163$ & $0.393$ & $-0.051$ & $0.443$ & $0.856$ & $+0.486\pm0.048$ & $+0.432\pm0.048$ & $+0.546\pm0.048$ & $0.4956$ & \mbox{no pref} \\
		GW240930\_234614 & $31.73$ & $23.95$ & $0.469$ & $0.548$ & $0.064$ & $0.889$ & $0.890$ & $-0.033\pm0.030$ & $-0.035\pm0.030$ & $-0.034\pm0.030$ & $0.3738$ & \mbox{no pref} \\
		GW241002\_030559 & $37.39$ & $29.87$ & $0.379$ & $0.440$ & $-0.113$ & $0.865$ & $0.876$ & $-0.085\pm0.020$ & $-0.078\pm0.020$ & $-0.063\pm0.019$ & $0.3622$ & \mbox{no pref} \\
		GW241006\_015333 & $24.65$ & $20.28$ & $0.322$ & $0.391$ & $0.074$ & $0.806$ & $0.856$ & $+0.167\pm0.036$ & $+0.133\pm0.036$ & $+0.206\pm0.036$ & $0.4200$ & \mbox{no pref} \\
		GW241007\_082943 & $38.48$ & $25.64$ & $0.616$ & $0.471$ & $0.052$ & $0.861$ & $0.880$ & $+0.200\pm0.026$ & $+0.136\pm0.026$ & $+0.174\pm0.026$ & $0.4275$ & \mbox{no pref} \\
		GW241009\_022835 & $36.53$ & $26.69$ & $0.425$ & $0.451$ & $0.078$ & $0.878$ & $0.882$ & $+0.048\pm0.023$ & $+0.048\pm0.023$ & $+0.063\pm0.023$ & $0.3924$ & \mbox{no pref} \\
		GW241009\_084816 & $12.71$ & $8.10$ & $0.416$ & $0.398$ & $-0.048$ & $0.863$ & $0.870$ & $+0.362\pm0.167$ & $+0.336\pm0.167$ & $+0.378\pm0.167$ & $0.4661$ & \mbox{no pref} \\
		GW241009\_220455 & $32.58$ & $25.67$ & $0.501$ & $0.480$ & $0.077$ & $0.881$ & $0.886$ & $+0.042\pm0.025$ & $+0.038\pm0.025$ & $+0.044\pm0.025$ & $0.3908$ & \mbox{no pref} \\
		GW241011\_233834 & $19.54$ & $5.96$ & $0.758$ & $0.354$ & $0.504$ & $0.138$ & $0.841$ & $-4.899\pm0.077$ & $-4.691\pm0.077$ & $-5.175\pm0.077$ & $0.0049$ & \mbox{pop} \\
		GW241101\_220523 & $40.12$ & $19.29$ & $0.352$ & $0.496$ & $0.113$ & $0.829$ & $0.888$ & $-0.117\pm0.032$ & $-0.091\pm0.032$ & $-0.099\pm0.032$ & $0.3553$ & \mbox{no pref} \\
		GW241102\_124058 & $10.91$ & $8.04$ & $0.218$ & $0.359$ & $0.064$ & $0.611$ & $0.850$ & $-2.206\pm0.152$ & $-2.259\pm0.152$ & $-2.168\pm0.152$ & $0.0667$ & \mbox{pop} \\
		GW241102\_144729 & $43.89$ & $31.55$ & $0.400$ & $0.434$ & $-0.118$ & $0.866$ & $0.885$ & $-0.100\pm0.017$ & $-0.023\pm0.017$ & $-0.079\pm0.017$ & $0.3588$ & \mbox{no pref} \\
		GW241109\_033317 & $42.85$ & $31.66$ & $0.424$ & $0.450$ & $-0.003$ & $0.880$ & $0.884$ & $-0.025\pm0.019$ & $+0.002\pm0.019$ & $-0.005\pm0.019$ & $0.3757$ & \mbox{no pref} \\
		GW241109\_115924 & $7.27$ & $5.14$ & $0.228$ & $0.344$ & $0.010$ & $0.732$ & $0.867$ & $+1.374\pm0.135$ & $+1.323\pm0.135$ & $+1.375\pm0.135$ & $0.6974$ & \mbox{acc} \\
		GW241110\_124123 & $16.11$ & $8.10$ & $0.631$ & $0.489$ & $-0.332$ & $0.733$ & $0.893$ & $-1.928\pm0.165$ & $-1.867\pm0.165$ & $-1.991\pm0.165$ & $0.0860$ & \mbox{pop} \\
		GW241111\_111552 & $25.21$ & $20.06$ & $0.363$ & $0.419$ & $0.082$ & $0.849$ & $0.875$ & $+0.310\pm0.025$ & $+0.281\pm0.025$ & $+0.342\pm0.025$ & $0.4536$ & \mbox{no pref} \\
		GW241113\_163507 & $19.26$ & $14.36$ & $0.805$ & $0.560$ & $0.496$ & $0.551$ & $0.883$ & $-3.774\pm0.624$ & $-3.528\pm0.624$ & $-3.996\pm0.624$ & $0.0148$ & \mbox{pop} \\
		GW241114\_024711 & $48.13$ & $24.79$ & $0.692$ & $0.475$ & $0.092$ & $0.771$ & $0.885$ & $+0.588\pm0.024$ & $+0.323\pm0.023$ & $+0.532\pm0.023$ & $0.5201$ & \mbox{acc weak} \\
		GW241114\_235258 & $11.26$ & $7.67$ & $0.320$ & $0.375$ & $0.056$ & $0.776$ & $0.867$ & $+0.602\pm0.150$ & $+0.561\pm0.150$ & $+0.626\pm0.150$ & $0.5234$ & \mbox{acc weak} \\
		GW241116\_151753 & $70.15$ & $24.35$ & $0.732$ & $0.517$ & $0.280$ & $0.849$ & $0.892$ & $+0.943\pm0.036$ & $+0.468\pm0.035$ & $+0.889\pm0.035$ & $0.6036$ & \mbox{acc weak} \\
		GW241124\_024914 & $41.94$ & $25.78$ & $0.362$ & $0.460$ & $0.020$ & $0.859$ & $0.884$ & $-0.152\pm0.024$ & $-0.083\pm0.024$ & $-0.129\pm0.024$ & $0.3475$ & \mbox{no pref} \\
		GW241125\_010116 & $60.22$ & $46.26$ & $0.406$ & $0.432$ & $0.064$ & $0.867$ & $0.881$ & $-0.296\pm0.018$ & $-0.075\pm0.017$ & $-0.277\pm0.018$ & $0.3166$ & \mbox{no pref} \\
		GW241127\_061008 & $58.15$ & $25.80$ & $0.606$ & $0.454$ & $-0.134$ & $0.489$ & $0.885$ & $+0.820\pm0.020$ & $+0.497\pm0.019$ & $+0.744\pm0.020$ & $0.5753$ & \mbox{acc weak} \\
		GW241129\_021832 & $30.15$ & $22.74$ & $0.370$ & $0.436$ & $0.131$ & $0.828$ & $0.871$ & $-0.004\pm0.026$ & $-0.023\pm0.026$ & $+0.018\pm0.026$ & $0.3805$ & \mbox{no pref} \\
		GW241130\_034908 & $30.19$ & $24.42$ & $0.312$ & $0.431$ & $0.049$ & $0.828$ & $0.885$ & $+0.226\pm0.024$ & $+0.193\pm0.024$ & $+0.262\pm0.024$ & $0.4337$ & \mbox{no pref} \\
		GW241130\_110422 & $11.10$ & $6.50$ & $0.264$ & $0.390$ & $0.039$ & $0.717$ & $0.882$ & $+0.879\pm0.153$ & $+0.836\pm0.153$ & $+0.905\pm0.153$ & $0.5891$ & \mbox{acc weak} \\
		GW241201\_055758 & $46.48$ & $32.86$ & $0.532$ & $0.491$ & $0.102$ & $0.883$ & $0.893$ & $+0.115\pm0.021$ & $+0.086\pm0.020$ & $+0.109\pm0.020$ & $0.4078$ & \mbox{no pref} \\
		GW241210\_060606 & $29.42$ & $21.26$ & $0.337$ & $0.423$ & $0.021$ & $0.836$ & $0.879$ & $+0.269\pm0.028$ & $+0.239\pm0.028$ & $+0.304\pm0.028$ & $0.4439$ & \mbox{no pref} \\
		GW241210\_120900 & $41.24$ & $23.45$ & $0.413$ & $0.463$ & $-0.069$ & $0.864$ & $0.887$ & $-0.037\pm0.027$ & $+0.011\pm0.027$ & $-0.023\pm0.027$ & $0.3730$ & \mbox{no pref} \\
		GW241225\_042553 & $12.45$ & $8.10$ & $0.448$ & $0.403$ & $0.116$ & $0.822$ & $0.854$ & $-3.994\pm0.056$ & $-4.005\pm0.056$ & $-3.990\pm0.056$ & $0.0119$ & \mbox{pop} \\
		GW241225\_082815 & $58.53$ & $36.27$ & $0.838$ & $0.477$ & $-0.225$ & $0.551$ & $0.893$ & $+1.357\pm0.020$ & $+0.305\pm0.018$ & $+1.208\pm0.020$ & $0.6939$ & \mbox{acc} \\
		GW241229\_155844 & $51.36$ & $33.17$ & $0.437$ & $0.474$ & $0.047$ & $0.878$ & $0.887$ & $-0.041\pm0.024$ & $+0.038\pm0.024$ & $-0.033\pm0.024$ & $0.3720$ & \mbox{no pref} \\
		GW241230\_084504 & $42.07$ & $33.86$ & $0.417$ & $0.462$ & $-0.150$ & $0.865$ & $0.881$ & $-0.104\pm0.018$ & $-0.060\pm0.018$ & $-0.095\pm0.018$ & $0.3580$ & \mbox{no pref} \\
		GW241230\_233618 & $67.96$ & $47.64$ & $0.499$ & $0.465$ & $-0.141$ & $0.884$ & $0.887$ & $-0.043\pm0.019$ & $+0.012\pm0.018$ & $-0.041\pm0.019$ & $0.3716$ & \mbox{no pref} \\
		GW241231\_054133 & $12.42$ & $7.09$ & $0.219$ & $0.422$ & $0.084$ & $0.577$ & $0.879$ & $+0.138\pm0.186$ & $+0.093\pm0.186$ & $+0.177\pm0.186$ & $0.4130$ & \mbox{no pref} \\
		GW250101\_011205 & $29.84$ & $21.36$ & $0.422$ & $0.453$ & $-0.049$ & $0.875$ & $0.884$ & $+0.109\pm0.031$ & $+0.094\pm0.032$ & $+0.125\pm0.031$ & $0.4063$ & \mbox{no pref} \\
		GW250104\_015122 & $45.87$ & $35.62$ & $0.400$ & $0.459$ & $0.080$ & $0.865$ & $0.885$ & $-0.078\pm0.018$ & $-0.029\pm0.017$ & $-0.064\pm0.017$ & $0.3638$ & \mbox{no pref} \\
		GW250108\_152221 & $55.70$ & $34.98$ & $0.462$ & $0.475$ & $0.015$ & $0.873$ & $0.888$ & $-0.039\pm0.019$ & $+0.032\pm0.018$ & $-0.035\pm0.019$ & $0.3725$ & \mbox{no pref} \\
		GW250109\_010541 & $36.72$ & $27.61$ & $0.556$ & $0.536$ & $-0.137$ & $0.890$ & $0.889$ & $-0.054\pm0.023$ & $-0.056\pm0.023$ & $-0.072\pm0.023$ & $0.3692$ & \mbox{no pref} \\
		GW250109\_074552 & $38.70$ & $30.18$ & $0.390$ & $0.410$ & $0.058$ & $0.856$ & $0.873$ & $+0.048\pm0.017$ & $+0.043\pm0.017$ & $+0.077\pm0.017$ & $0.3923$ & \mbox{no pref} \\
		GW250114\_082203 & $33.77$ & $32.21$ & $0.075$ & $0.084$ & $-0.023$ & $0.246$ & $0.263$ & $+0.492\pm0.011$ & $+0.406\pm0.011$ & $+0.593\pm0.011$ & $0.4972$ & \mbox{no pref} \\
		GW250116\_015318 & $34.21$ & $21.33$ & $0.511$ & $0.481$ & $0.152$ & $0.881$ & $0.889$ & $+0.097\pm0.043$ & $+0.078\pm0.043$ & $+0.092\pm0.043$ & $0.4035$ & \mbox{no pref} \\
		GW250118\_023225 & $43.76$ & $30.46$ & $0.413$ & $0.471$ & $0.049$ & $0.877$ & $0.887$ & $-0.086\pm0.020$ & $-0.037\pm0.020$ & $-0.072\pm0.020$ & $0.3620$ & \mbox{no pref} \\
		GW250118\_055802 & $10.27$ & $6.94$ & $0.399$ & $0.424$ & $0.113$ & $0.801$ & $0.877$ & $-0.739\pm0.387$ & $-0.760\pm0.387$ & $-0.732\pm0.387$ & $0.2319$ & \mbox{pop} \\
		GW250118\_170523 & $38.42$ & $30.10$ & $0.434$ & $0.519$ & $0.126$ & $0.880$ & $0.886$ & $+0.006\pm0.020$ & $-0.014\pm0.020$ & $+0.006\pm0.019$ & $0.3826$ & \mbox{no pref} \\
		GW250119\_025138 & $33.66$ & $27.05$ & $0.384$ & $0.390$ & $0.054$ & $0.865$ & $0.870$ & $+0.158\pm0.020$ & $+0.129\pm0.021$ & $+0.191\pm0.020$ & $0.4179$ & \mbox{no pref} \\
		GW250119\_190238 & $11.54$ & $9.95$ & $0.199$ & $0.271$ & $0.089$ & $0.572$ & $0.708$ & $-2.647\pm0.035$ & $-2.716\pm0.035$ & $-2.575\pm0.035$ & $0.0442$ & \mbox{pop} \\
	\end{longtable}

\end{document}